\documentclass{article}

\usepackage[utf8]{inputenc}
\usepackage{amsmath}
\usepackage{caption}
\usepackage{subcaption}
\usepackage{graphicx}
\usepackage{xcolor}
\usepackage{hyperref}
\usepackage{biblatex} 
\usepackage{authblk}

\begin{document}

\title{Photonic Extreme Learning Machine based on frequency multiplexing}
\date{}
\author[1]{Alessandro Lupo\footnote{alessandro.lupo@ulb.be}}
\author[1]{Lorenz Butschek}
\author[1]{Serge Massar\footnote{serge.massar@ulb.be}}

\affil[1]{Laboratoire d’Information Quantique, CP 224, Universitè libre de Bruxelles, Av. F. D. Roosevelt 50, B-1050,
Bruxelles, Belgium}


\maketitle

\begin{abstract}
The optical domain is a promising field for physical implementation of neural networks, due to the speed and parallelism of optics. Extreme Learning Machines (ELMs) are feed-forward neural networks in which only output weights are trained, while internal connections are randomly selected and left untrained. Here we report on a photonic ELM based on a frequency-multiplexed fiber setup. Multiplication by output weights can be performed either offline on a computer, or optically by a programmable spectral filter. We present both numerical simulations and experimental results on classification tasks and a nonlinear channel equalization task.
\end{abstract}

\section{Introduction}\label{sec:introduction}

\begin{figure}[ht]
\begin{subfigure}{.5\textwidth}
  \centering
  \includegraphics[width=.8\linewidth]{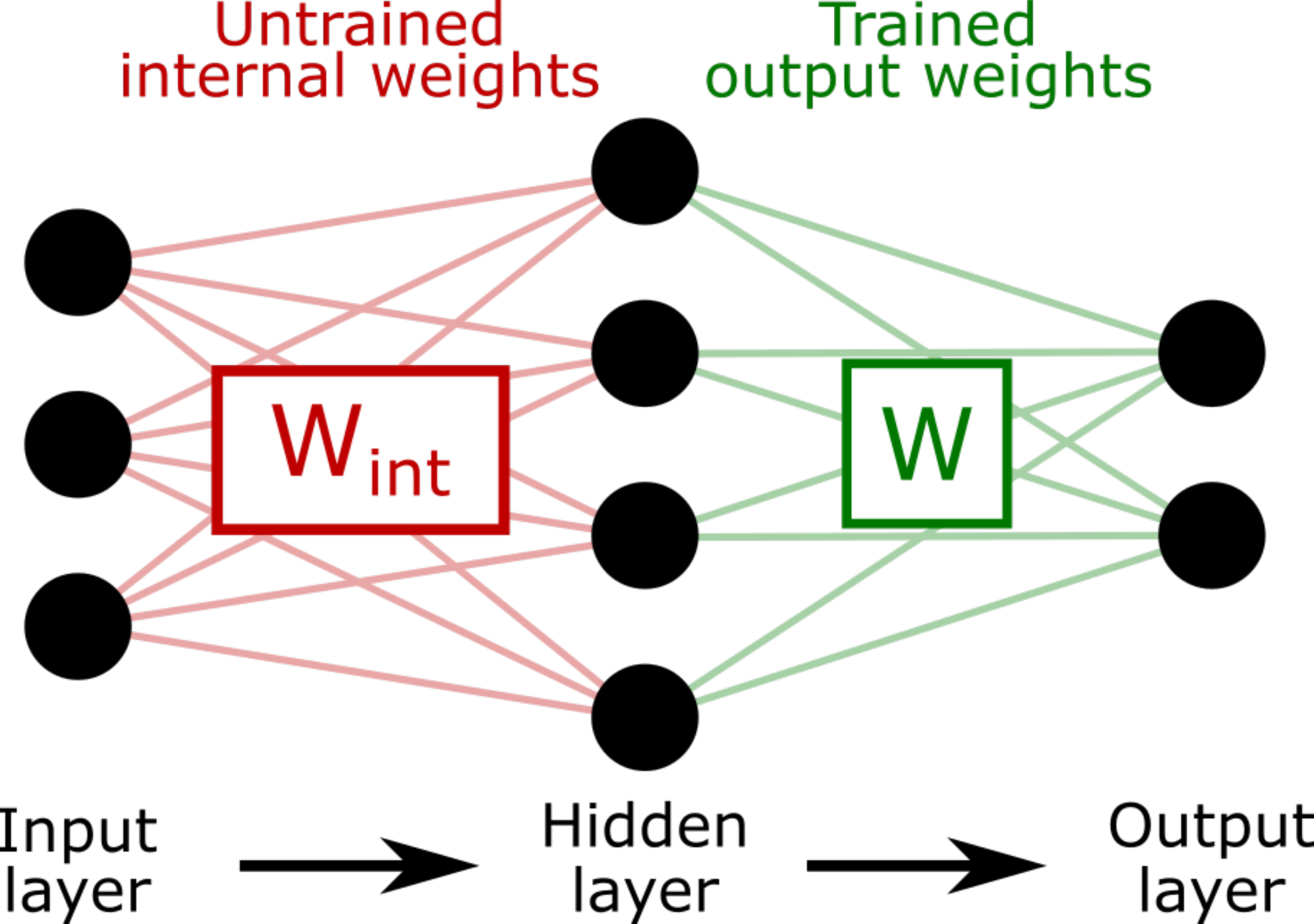}  
  \caption{Single hidden layer ELM}
  \label{fig:ELM_net}
\end{subfigure}
\begin{subfigure}{.5\textwidth}
  \centering
  \includegraphics[width=.8\linewidth]{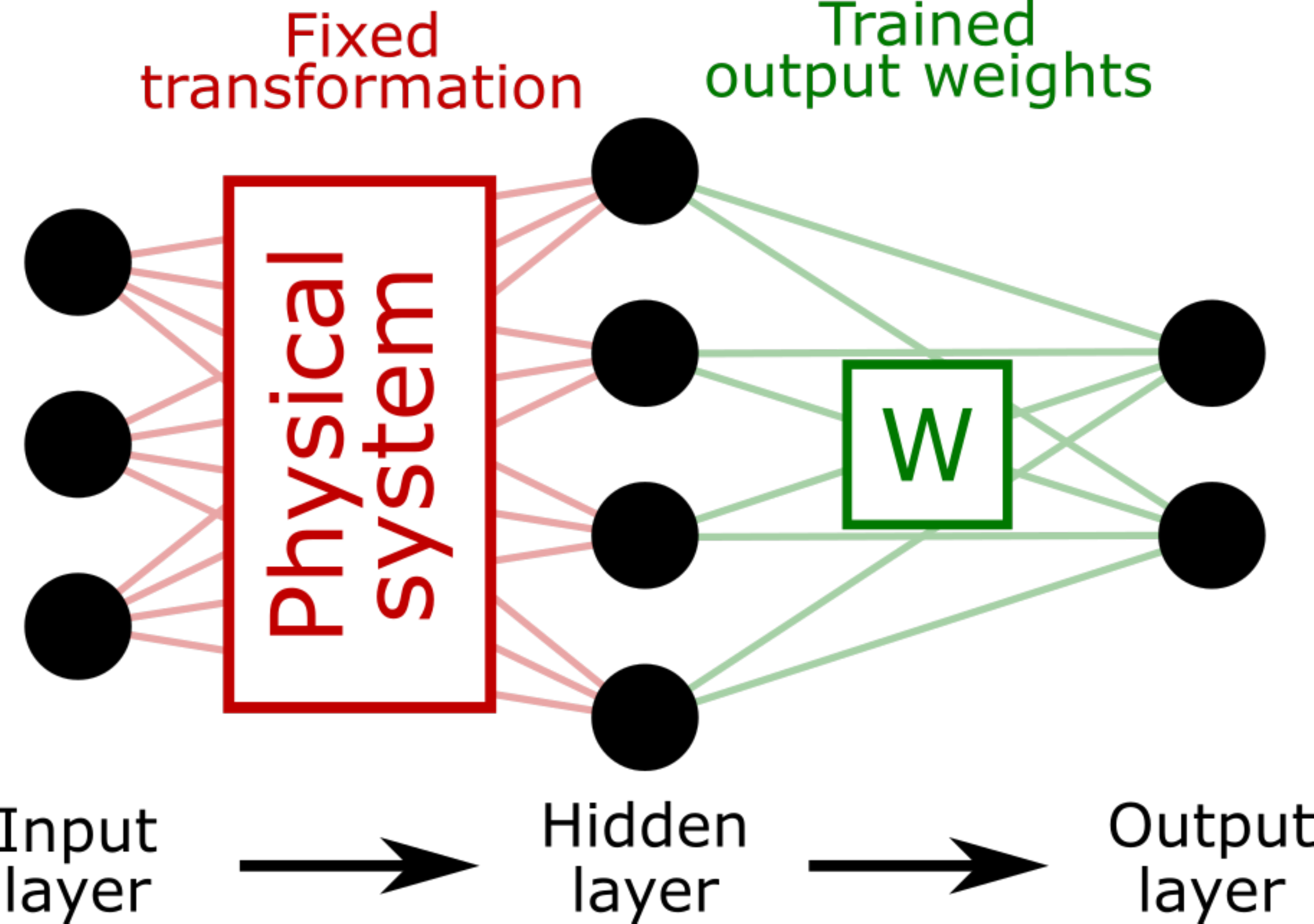}  
  \caption{Physical ELM}
  \label{fig:ELM_phys}
\end{subfigure}
\caption{An ELM (a) is trained by acting on the output weights, $W$, in green, while the weights between input and hidden layer, $W_{int}$, in red, are randomly selected and kept fixed. In a physical implementation of an ELM (b) the untrained connections between input and hidden layer are substituted with the action of a physical system; the output of this system constitutes the hidden layer of the network.}
\label{fig:ELM}
\end{figure}

Feed-Forward Neural networks (FFNs) are among the most employed machine learning algorithms due to their simplicity and their universal approximation property. The training procedure of FFNs is usually both time and power consuming, consisting in optimizing each network weight via slow gradient descent algorithms. Extreme Learning Machines (ELMs, Figure \ref{fig:ELM_net}) are FFNs composed of a single hidden layer in which only the output weights are updated, usually in a single step, while other parameters remain fixed during training, thereby speeding up the learning \cite{huang2004, huang2006extreme, huang2015extreme, gonon2020approximation}.

The ELM paradigm can be implemented in physical systems of various natures. The transformation that the chosen system performs between its input space and output space is analogous to the untrained set of internal connections of the ELM (Figure \ref{fig:ELM_phys}). A system that maps its input space into a higher dimensional output space through a nonlinear transformation is expected to be a good candidate for an ELM. The training of such a “physical ELM” consists in the search for the optimal linear transformation which, acting on the system output, best approximates the desired target. The coefficients of such a linear transformation are analogous to the output weights of the network. 

The optical domain offers good parallelization capabilities, many nonlinearities and high speed, thus it is considered a promising substrate for neural network implementations \cite{xu2021survey}. Many schemes used to implement optical neural networks can also be used to implement ELMs. For instance, free space propagation through scattering media, exploited by Diffractive Deep Neural Networks ($\textrm{D}^2\textrm{NN}$) \cite{Lin2018, Zhou2021}, has been employed for ELMs \cite{saade2016}; similarly, time-multiplexed fiber loops, extensively exploited for Reservoir Computing (RC) \cite{Appeltant2011, Paquot2012}, have been employed for ELMs \cite{ortin2015unified}. In these ELM implementations, the physical system output is recorded by a computer and the final transformation, i.e.\ the multiplication by output weights, is calculated digitally.

\begin{figure}[ht]
  \centering
  \includegraphics[width=.65\linewidth]{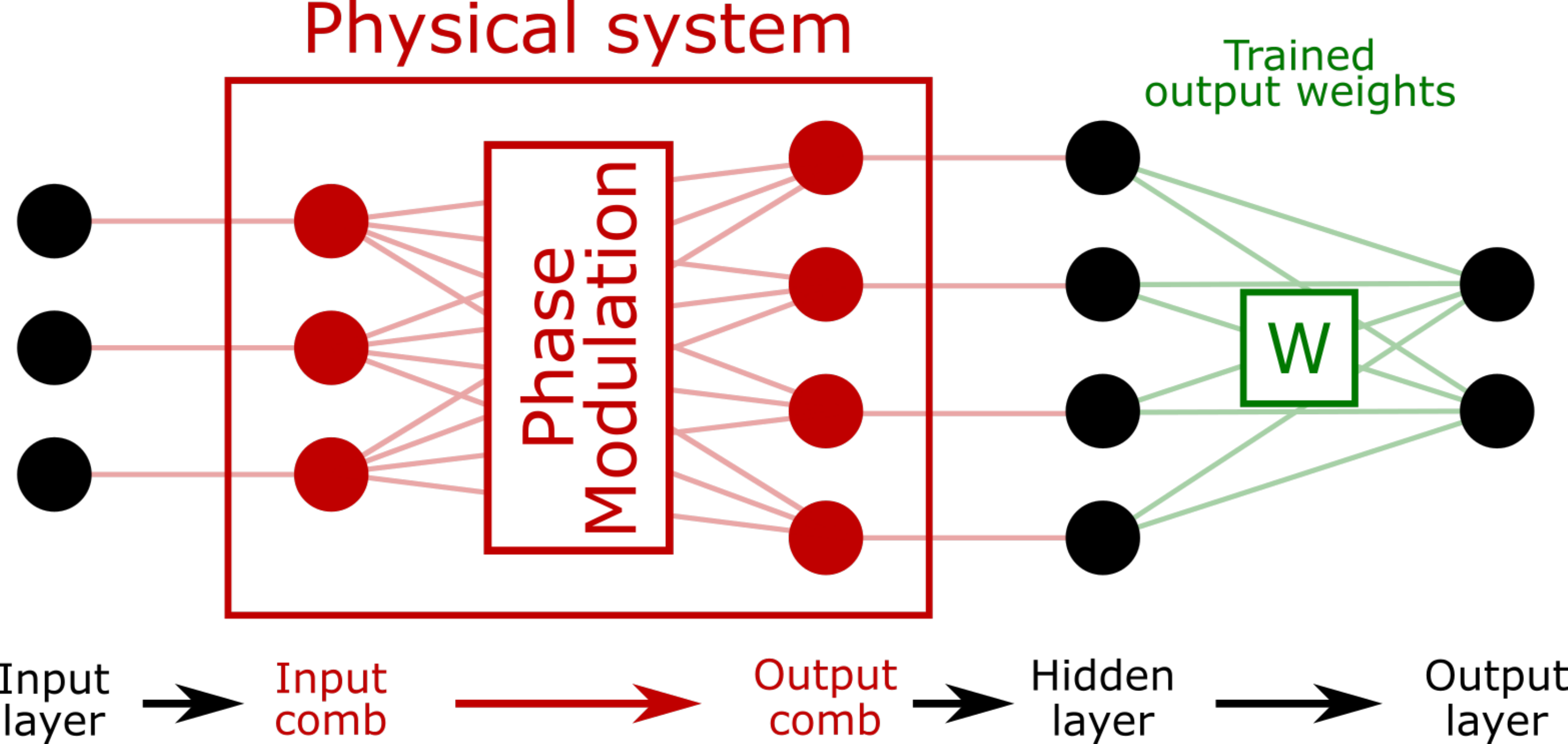}  
\caption{Conceptual scheme of the ELM based on frequency multiplexing. The physical system, in red, generates a frequency comb (input comb) and subsequently transforms it into a new comb (output comb), mixing its frequencies through phase modulation. The input layer is encoded in the input comb, hence the output comb plays the role of the hidden layer. Each hidden node is a linear combination of input nodes. The only nonlinearity is the quadratic one realised by the readout photodiodes.}
\label{fig:ELM_simp}
\end{figure}

Here we present a photonic ELM based on frequency multiplexing (Figure \ref{fig:ELM_simp}), where information processing is mostly performed optically, including the multiplication by output weights. The states of both input and hidden nodes are encoded in the amplitudes of different lines of a frequency comb. The comb is generated by a Phase Modulator acting on monochromatic laser light. Input features are encoded in the amplitudes of the comb by a programmable spectral filter. The input layer is transformed into the hidden layer via frequency mixing carried out by a second Phase Modulator: this technique, introduced in Quantum Optics \cite{Merolla1999, Olislager2010}, has been already employed for optical Reservoir Computing \cite{butschek2020parallel}. A second programmable spectral filter is used either to apply output weights, thus optically generating the output layer, or to scan the frequencies of the hidden layer comb, thus measuring the state of each hidden node. The only nonlinearity is a quadratic nonlinearity performed by the readout photodiodes.

In section \ref{sec:experimental_setup} we describe the experimental setup and the model employed in numerical simulation. In section \ref{sec:methods} we describe all the experiment phases, from input to performance evaluation, including the training algorithm and the optical weighting scheme. In Section \ref{sec:results} we describe the results obtained on different classification tasks and on Nonlinear Channel Equalization task, discussing their comparison with simulations, other machine learning algorithms and previous literature. We also discuss the dependence of performances on hyperparameters. Section \ref{sec:conclusion} contains conclusions and perspectives.

\section{Experimental system}\label{sec:experimental_setup}
\subsection{Experimental setup}
\begin{figure}[ht]
  \centering
  \includegraphics[width=\textwidth]{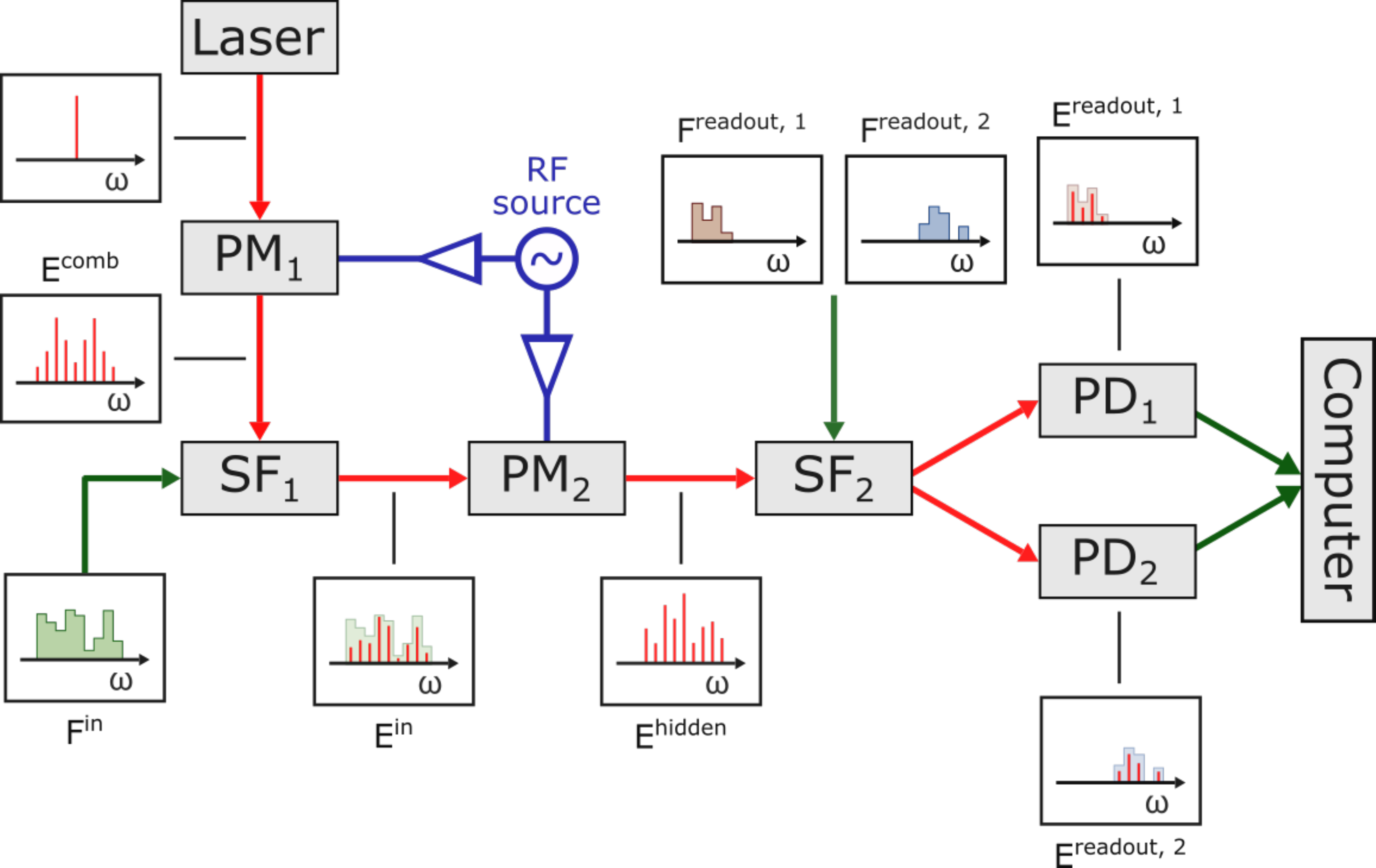}  
  \caption{Scheme of the experimental setup. Red lines represent optical connections, green lines represent input from the computer and blue lines represent RF connections. The first Phase Modulator, $\textrm{PM}_1$, generates a frequency comb out of monochromatic laser radiation. The first programmable spectral filter, $\textrm{SF}_1$, encodes input features in this comb, thus generating the input layer. The second Phase Modulator, $\textrm{PM}_2$, mixes the input comb components generating the hidden layer. The second programmable spectral filter, $\textrm{SF}_2$, is employed for the readout. The two photodiodes $\textrm{PD}_1$ and $\textrm{PD}_2$ provide an integrated reading of all the optical power impinging on them. A computer drives the programmable filters (connections not shown) and records the photodiodes measurements.}
  \label{fig:setup}

\end{figure}

Our experimental setup is depicted in Figure \ref{fig:setup}. The light source is a C-Band continuous wave laser propagating in polarization-maintaining fibers. The two Phase Modulators, $\textrm{PM}_1$ and $\textrm{PM}_2$, are driven by the same Radio Frequency (RF) signal generator at frequency ${\Omega/2\pi = 16.96860\textrm{ GHz}}$. $\Omega$ defines the spacing of the comb, as shown in Section \ref{sec:electric_field}, and its exact value is not important. The same RF signal goes through two amplifiers which provide two different fixed gains (hence the RF powers reaching the two PMs cannot be set independently, as only the RF generator power can be tuned). During the experiment, $\textrm{PM}_1$ and $\textrm{PM}_2$ are driven by RF powers of $30\textrm{ dBm}$ and $20\textrm{ dBm}$ respectively. The strength of modulation is better characterized by the dimensionless number $m = \frac{\pi V}{V_\pi}$, where $V$ is the amplitude of the signal applied to the $\textrm{PM}$, and $V_\pi$ is the $\textrm{PM}$ characteristic voltage. In our setup, $m_1\approx7.87$ and $m_2\approx2.18$. The programmable spectral filters $\textrm{SF}_1$ and $\textrm{SF}_2$ are two Finisar Waveshapers, model $1000$ and $4000$ respectively. $\textrm{SF}_1$ is employed to encode the input, applying the proper attenuation to each component of the comb. $\textrm{SF}_2$, instead, allows to apply two different filters, redirecting the two results to two different outputs. The time to set a new spectral filter is approximately $500\textrm{ ms}$. The two outputs of $\textrm{SF}_2$ are connected to two photodiodes, $\textrm{PD}_1$ and  $\textrm{PD}_2$, and their readings are transferred to a computer. Each hidden node can be read by using $\textrm{SF}_2$ to implement the corresponding notch filter. Since two filters can be set simultaneously, up to two different nodes can be read at the same time. To perform optical multiplication with output weights, instead, more complex filter shapes are set in $\textrm{SF}_2$, in such a way that each photodiode, integrating the optical power over the whole spectrum, measures a specific linear combination of comb component powers.
The programmable filter $\textrm{SF}_1$ provides a $20\textrm{ GHz}$ bandwidth resolution, while $\textrm{SF}_2$ provides a $10\textrm{ GHz}$ resolution. Considering the value of $\Omega$, equal to the spacing between comb lines, these filter resolutions should in principle be enough to fix the attenuation of each comb component separately. However, we measured a slight crosstalk effect between two adjacent lines filtered by $\textrm{SF}_1$, meaning that the value encoded on one input node may slightly influence the adjacent ones. Simulations suggest that this crosstalk has no effect on performances, but could be avoided by increasing $\Omega$ or choosing a better resolved programmable spectral filter.

\subsection{Description of the electric field}
\label{sec:electric_field}
A Phase Modulator acts on monochromatic laser radiation as follows:
\begin{equation}
\label{eq:jacobi-anger}
E_0e^{-i\omega t}\rightarrow E_0e^{-i\omega t}e^{-im\cos{(\Omega t)}}=E_0e^{-i\omega t}\sum_{k=-\infty}^{+\infty}i^kJ_k(m)e^{-ik\Omega t},
\end{equation}
where $E_0$ is the input electric field amplitude, $\omega$ is the input electric field angular frequency, $\Omega$ is the RF frequency driving the $\textrm{PM}$, $m$ is its modulation strength, and $J_\nu(m)$ represent the Bessel functions of first kind. The series expansion of the term $e^{-im\cos{(\Omega t)}}$ is known as Jacobi-Anger expansion. The coefficients of this expansion decrease when $|k|$ increases, thus the series can be truncated in numerical simulations.

We define $E^\textrm{comb}$ the electric field at the output of $\textrm{PM}_1$; $E^\textrm{in}$ the electric field at the output of $\textrm{SF}_1$; $E^\textrm{hidden}$ the electric field at the output of $\textrm{PM}_2$ and $E^\textrm{readout, 1}$ and $E^\textrm{readout, 2}$ the two electric fields at the two outputs of $\textrm{SF}_2$, hence at the inputs of $\textrm{PD}_1$ and $\textrm{PD}_2$ (see Figure \ref{fig:setup}). These definitions reflect the function of the fields in the ELM context: $E^\textrm{comb}$ represents the blank comb before any input is encoded on it, $E^\textrm{in}$ represents the input layer of the ELM and $E^\textrm{hidden}$ represents the hidden layer. Note that $E^\textrm{readout, 1}$ and $E^\textrm{readout, 2}$ do not represent necessarily the output layer of the ELM: their content depends on how $\textrm{SF}_2$ is set, as described in Section \ref{sec:readout}, and needs postprocessing to reconstruct the actual output layer. The shape of the programmable spectral filters are described by the attenuations that they apply on the frequencies $\omega+k\Omega$, which are the frequencies of the comb components, i.e.\ the frequencies of each node. We define $F_k^\textrm{in}$ the attenuation that the filter set on $\textrm{SF}_1$ applies to the frequency $\omega+k\Omega$, and $F_k^\textrm{readout, 1}$ and $F_k^\textrm{readout, 2}$ the attenuations that the two filters set on $\textrm{SF}_2$ apply to the frequency $\omega+k\Omega$.
Hence, the electric fields across the setup are described by the following equations:
\begin{align}
    E^\textrm{comb}(t)   &= \sum_kE_k^\textrm{comb}e^{-i(\omega+k\Omega)t}, &E_k^\textrm{comb} &= E_0i^kJ_k(m_1);\label{eq:Ecomb}\\
    E^\textrm{in}(t)  &= \sum_kE_k^\textrm{in}e^{-i(\omega+k\Omega)t}, &E_k^\textrm{in} &= E_k^\textrm{comb}\sqrt{F_k^\textrm{in}};\label{eq:Einput}\\
    E^\textrm{hidden}(t) &= \sum_kE_k^\textrm{hidden}e^{-i(\omega+k\Omega)t}, &E_k^\textrm{hidden} &= \sum_pE^\textrm{in}_pi^{k-p}J_{k-p}(m_2);\label{eq:Ehidden}\\
    E^\textrm{readout, 1}(t) &= \sum_kE_k^\textrm{readout, 1}e^{-i(\omega+k\Omega)t}, &E_k^\textrm{readout, 1} &= E_k^\textrm{hidden}\sqrt{F_k^\textrm{readout, 1}};\label{eq:Ereadout1}\\
    E^\textrm{readout, 2}(t) &= \sum_kE_k^\textrm{readout, 2}e^{-i(\omega+k\Omega)t}, &E_k^\textrm{readout, 2} &= E_k^\textrm{hidden}\sqrt{F_k^\textrm{readout, 2}}\label{eq:Ereadout2}.
\end{align}
The photodiodes $\textrm{PD}_1$ and $\textrm{PD}_2$ provide measurements of the overall optical intensity integrated over the whole spectral extension of the filtered comb:
\begin{align}
I^\textrm{readout, 1} &= \left|<E^\textrm{readout, 1}(t)>\right|^2 = \sum_k  \left| E_k^\textrm{readout, 1}\right|^2,\\
I^\textrm{readout, 2} &= \left|<E^\textrm{readout, 2}(t)>\right|^2 = \sum_k  \left| E_k^\textrm{readout, 2}\right|^2,
\end{align}
where $<\cdot>$ indicates a time average.

\begin{figure}[ht]
\begin{subfigure}{.32\textwidth}
  \centering
  \includegraphics[width=.99\linewidth]{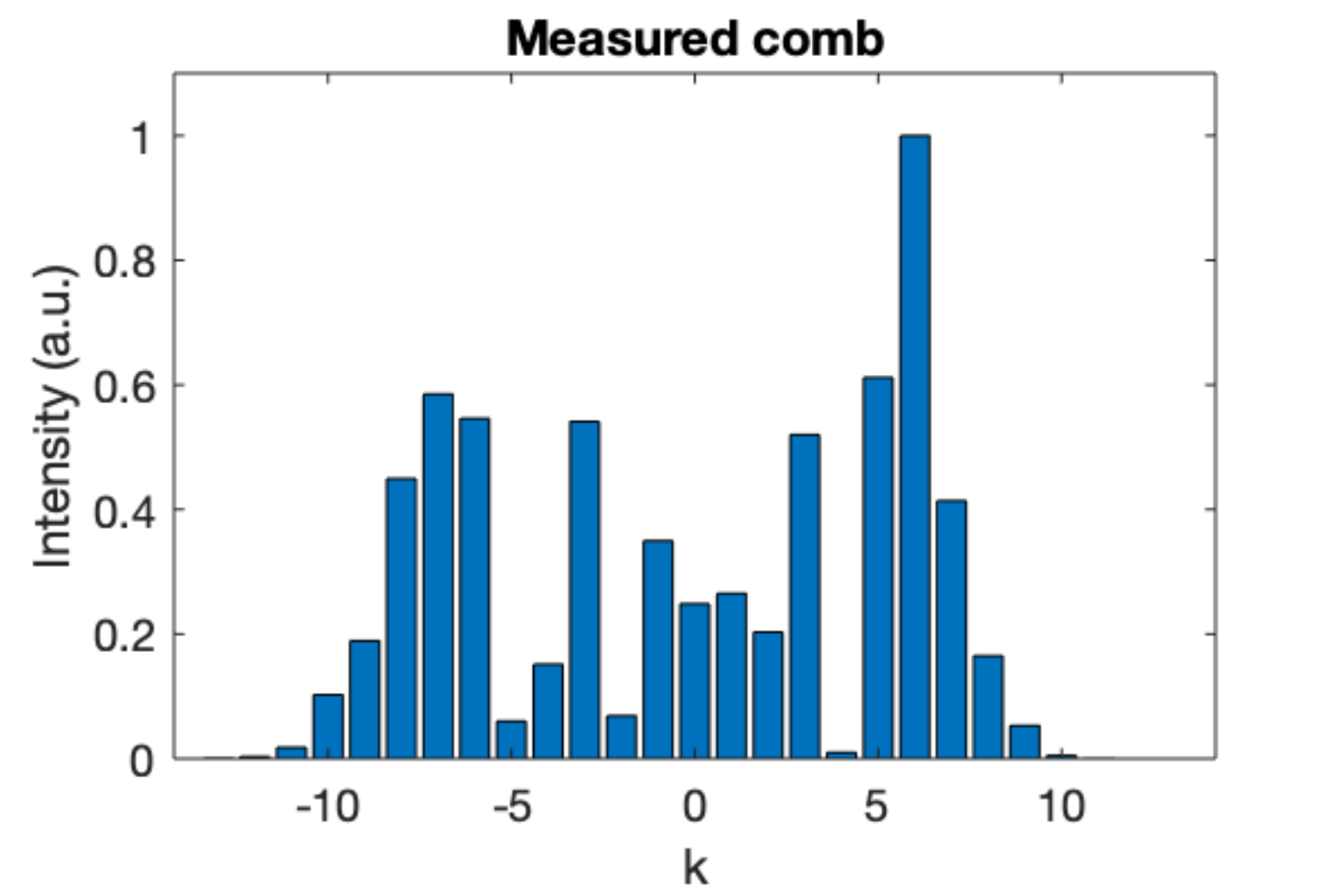}  
  \caption{}
  \label{fig:measured_comb}
\end{subfigure}
\begin{subfigure}{.32\textwidth}
  \centering
  \includegraphics[width=.99\linewidth]{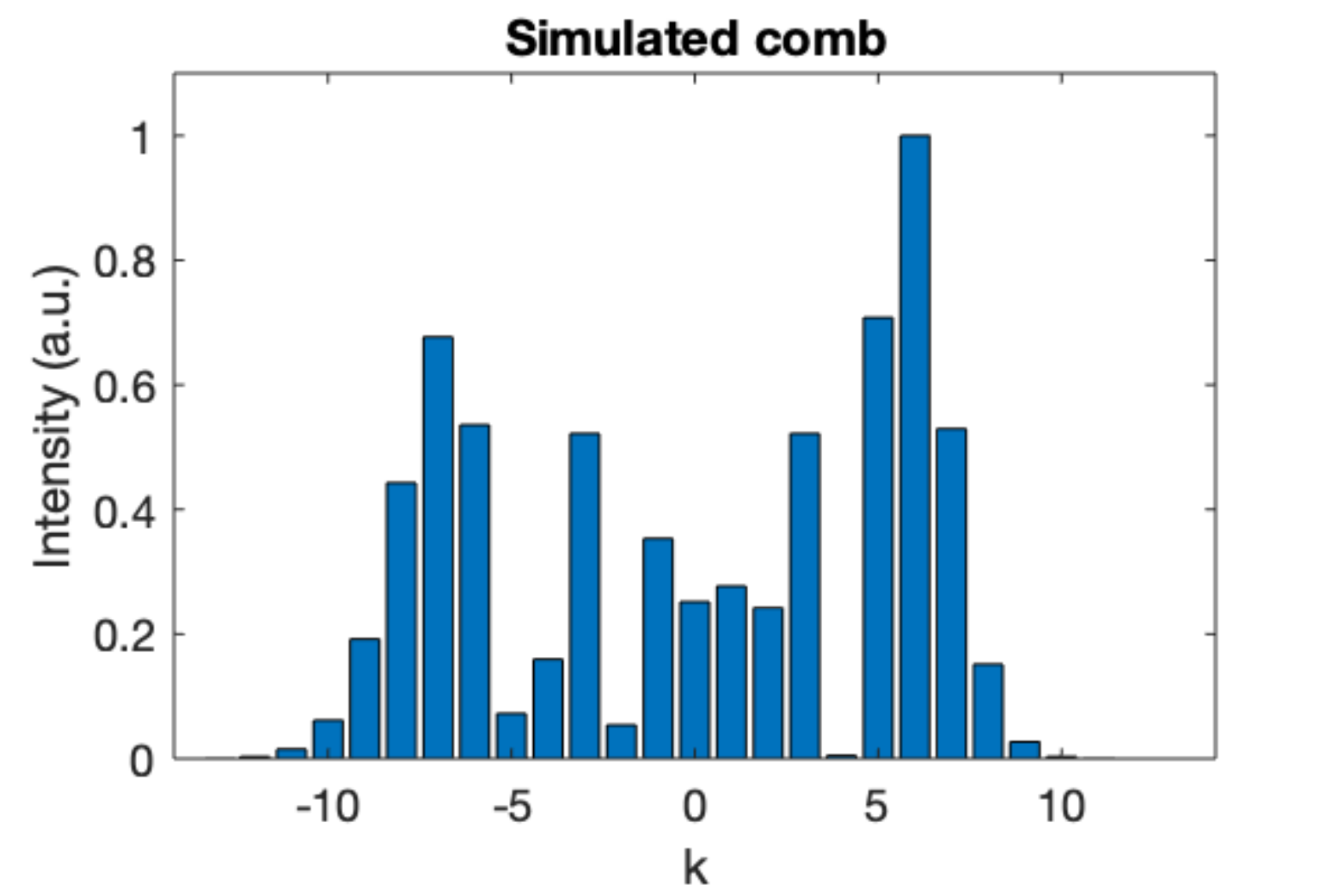}  
  \caption{}
  \label{fig:simulated_comb}
\end{subfigure}
\begin{subfigure}{.32\textwidth}
  \centering
  \includegraphics[width=.99\linewidth]{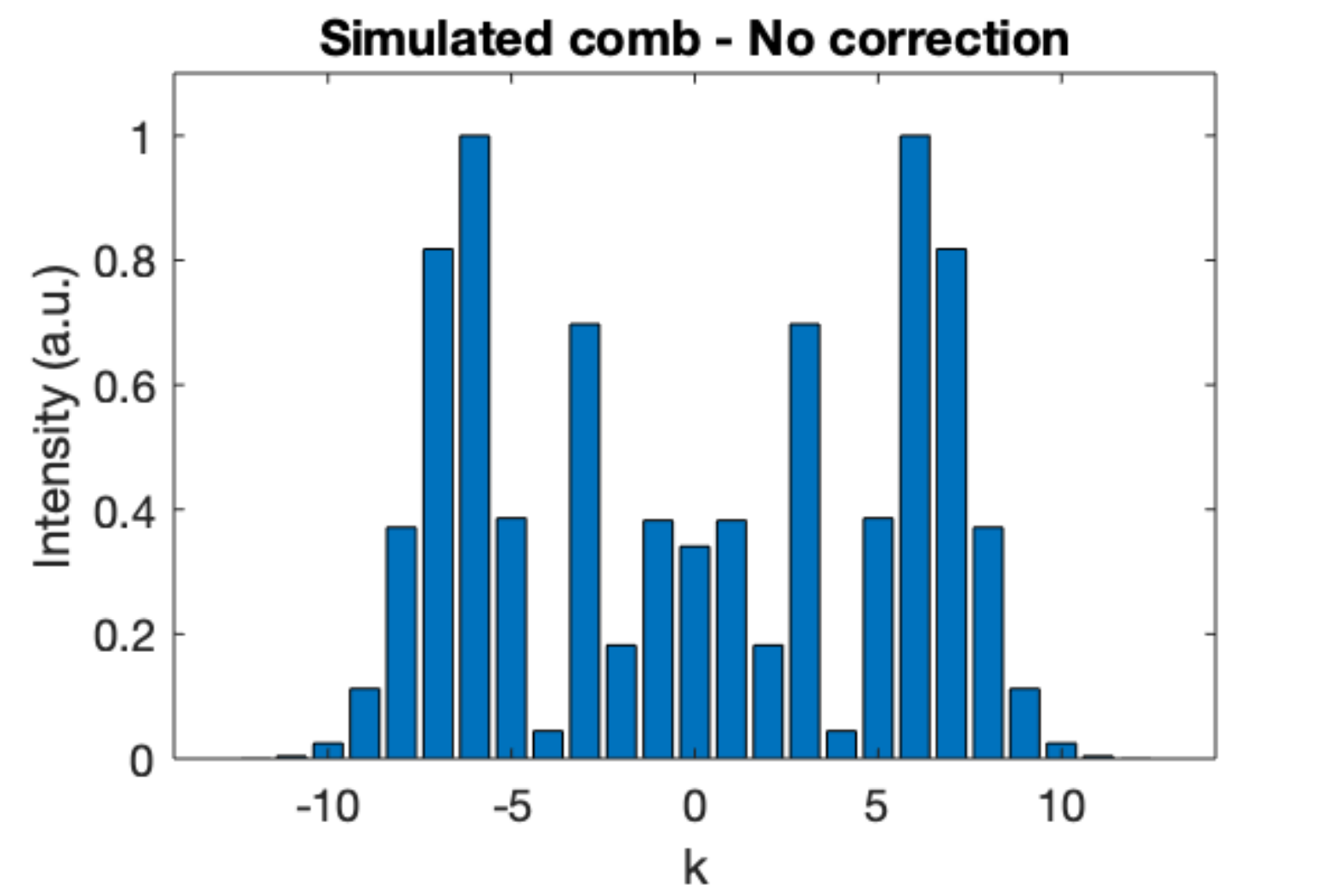}  
  \caption{}
  \label{fig:simulated_comb_no_epsilon}
\end{subfigure}
\caption{Comb intensities after $\textrm{PM}_1$ corresponding to the parameters reported in the text. Measurements (a); simulation accounting for the second harmonic correction (b); simulation without correction (c).}
\label{fig:comb}
\end{figure}

The model described by Eq. \eqref{eq:jacobi-anger} generates symmetrical input combs (Figure \ref{fig:simulated_comb_no_epsilon}), while the comb measured experimentally shows clear asymmetries (Figures \ref{fig:measured_comb}). The asymmetry suggests the presence of a second harmonic of the RF signal driving the Phase Modulators. In order to achieve realistic simulations, we correct Eq. \eqref{eq:jacobi-anger} as follows:
\begin{equation}
\label{eq:second_harmonic}
E_0e^{-i\omega t}\rightarrow E_0e^{-i\omega t}e^{-im\cos{(\Omega t)}}e^{-i\epsilon m\cos{(2\Omega t+\Phi)}},
\end{equation}
where the second exponential factor accounts for a new second harmonic effect and $\epsilon$ represents its strength. Simulations can still be performed easily, since the two exponential factors featuring $\cos{(\Omega t)}$ and $\cos{(2\Omega t+\Phi)}$ can be expanded in two Jacobi-Anger series, as follows:
\begin{equation}
\label{eq:correction_sh}
e^{-im\cos{(\Omega t)}-i\epsilon m\cos{(2\Omega t+\Phi)}} = \sum_{k=-\infty}^{+\infty}\sum_{p=-\infty}^{+\infty} i^{k+p}J_k(m)J_p(\epsilon m)e^{-i(k+2p)\Omega t-ip\Phi},
\end{equation}
where, as before, the sums can be truncated when the coefficients get small enough. After manipulating the indexes in Eq. \eqref{eq:correction_sh}, we can correct Eq. \eqref{eq:Ecomb} accounting for comb asymmetries:
\begin{equation}
    E_k^\textrm{comb} = E_0 \sum_{p=-\infty}^{+\infty}i^{k-p}J_{k-2p}(m_1)J_p(\epsilon m_1)e^{-ip\Phi}.
\end{equation}
The values of $\epsilon$ and $\Phi$ have been fitted to match experimental measures of the combs generated by $\textrm{PM}_1$. We found $\epsilon=0.0471$ and $\Phi=1.31\textrm{ rad}$.
The new equation provide a more realistic comb, as shown in Figure \ref{fig:simulated_comb}.

A similar expression can be derived for Eq. \eqref{eq:Ehidden}, but seems not to be necessary, since $\textrm{PM}_2$ is driven by a weaker RF signal and exhibits weaker nonlinearity. 

\begin{figure}[ht]
\centering
\includegraphics[width=.99\linewidth]{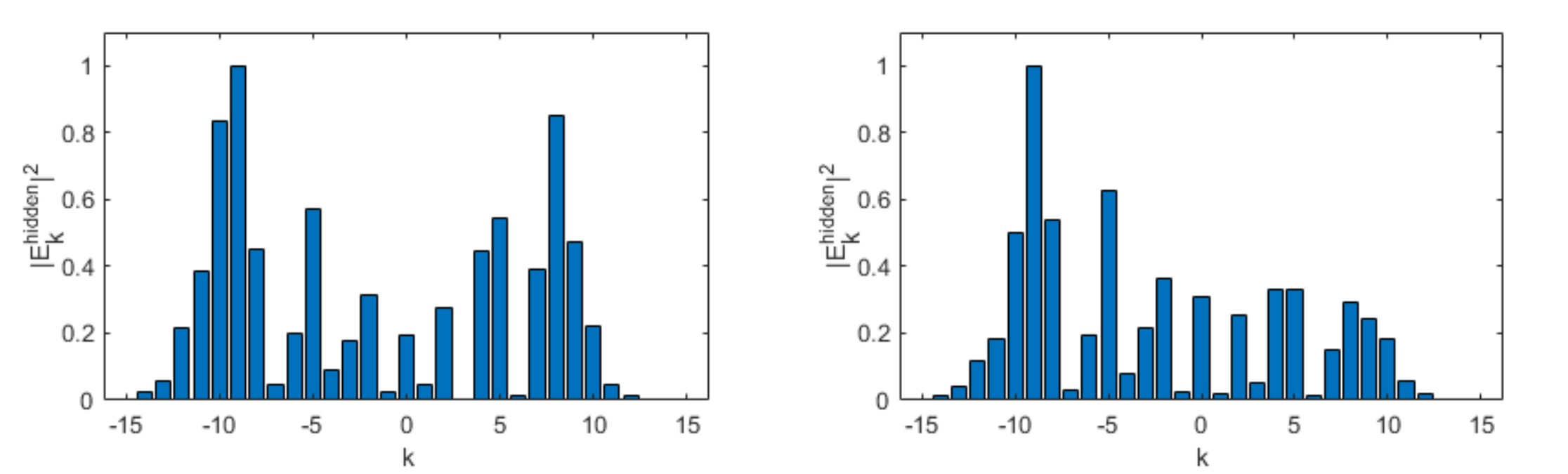}  
\caption{Two hidden combs generated from two different input layers during experiment. Note that the comb contains about $31$ lines. $k=0$ correspond to the frequency of the laser source.}
\label{fig:hidden_layer}
\end{figure}

Regardless of the number of input nodes, the hidden layer is always considered to be composed of $31$ nodes, i.e.\ only the 31 most central lines of the comb are read and linearly combined. This is because, given the values of $m_1$ and $m_2$, components $E^\textrm{hidden}_k$ with $|k|>15$ are always too weak to encode information, as shown in Figure \ref{fig:hidden_layer}. Note that $E^\textrm{hidden}_k$ may be too weak to be measured even for certain $k\in[-15,\,15]$, but these "silent" nodes are not expected to affect our training algorithm (see Section \ref{sec:training}).

\section{Principle of operation}\label{sec:methods}
\subsection{Notation}
In the following we indicate with $\textbf{u}$ a single set of input features supplied to the ELM, i.e.\ a single input layer; with $\mathbf{\tilde{y}}$ its corresponding target output layer, i.e.\ the correct output expected from a well trained network; and with $\textbf{h}$ the hidden layer of the ELM. The output layer $\mathbf{y}$ is generated multiplying the hidden layer $\textbf{h}$ with the set of output weights $\mathbf{W}$: $\mathbf{y}=\mathbf{h}\cdot\mathbf{W}$. The multiplication can be performed digitally or optically, as described in the following. $\textbf{u}$ and $\textbf{h}$ are row vectors; $\mathbf{{y}}$ and $\mathbf{\tilde{y}}$ are scalars if the task require only one output node, or row vectors otherwise; $\mathbf{W}$ is a column vector if the output layer contains only one node, or a matrix otherwise. To describe the training phase, it is useful to include all the input layers submitted to the network, all the corresponding hidden layers and all the corresponding target outputs in matrices. We hence define $\mathbf{U}$, $\mathbf{H}$ and $\mathbf{\tilde{Y}}$ in such a way that the i-th row of $\mathbf{U}$ represents the i-th set of input features submitted to the network, the i-th row of $\mathbf{H}$ represents the corresponding hidden layer generated by the network, and the i-th row of $\mathbf{\tilde{Y}}$ represents the corresponding target output layer. $\mathbf{U}$ and $\mathbf{H}$ are matrices, while $\mathbf{\tilde{Y}}$ is a column vector if the task requires only one output node, or a matrix otherwise.

\subsection{Training algorithm}\label{sec:training}
The training consists in finding the optimal set of output weights, $\mathbf{W}$, such that when the input $\textbf{u}$ is presented to the network, the output layer approximates the corresponding target $\mathbf{\tilde{y}}$, i.e.: $\mathbf{y}=\textbf{h}\cdot\textbf{W}\approx\mathbf{\tilde{y}}$.
Note that only the output weights $\mathbf{W}$ are trained, while the internal mechanism which transforms the input layer into the hidden layer is left untouched. Hence the ELM training consists in a single operation and does not require slow gradient descent algorithms. In this work we employ ridge regression algorithm to estimate the optimal set of output weights $\textbf{W}$. Ridge regression consists in the minimization of the quantity
\begin{equation}
\label{eq:ridge}
||\mathbf{H}\mathbf{W}-\mathbf{\tilde{Y}}||^2+||\lambda\mathbf{W}||^2,
\end{equation}
where $\lambda$ is a regularization parameter whose purpose is described below. The $\mathbf{W}$ minimizing \eqref{eq:ridge} is:
\begin{equation}
\mathbf{W} =(\mathbf{H}^T\mathbf{H}+\lambda^2)^{-1}\mathbf{H}^T\mathbf{\tilde{Y}}, 
\end{equation}
where $\left(\cdot\right)^T$ indicates the transposed matrix and $\left(\cdot\right)^{-1}$ the inverse matrix. In our system no hidden node will measure exactly zero, and the algorithm may erroneously attribute importance to dark noise, setting enormous weight to silent hidden nodes. The regularization parameter $\lambda$ defines a penalty for having high components in the vector $\mathbf{W}$, thus preventing this error. The optimal value for $\lambda$ depends on the task and is obtained by testing different possibilities. It is worth introducing here also the Ordinary Least Squares (OLS) estimation, which is equivalent to ridge regression with $\lambda=0$. As described in Section \ref{sec:readout}, OLS is employed during the optical weighting. The solution in this case is
\begin{equation}
\label{eq:least_squares}
\mathbf{W}=\textrm{pinv}(\mathbf{H})\mathbf{\tilde{Y}},
\end{equation}
where $\textrm{pinv}(\mathbf{H})=(\mathbf{H}^T\mathbf{H})^{-1}\mathbf{H}^T$ is the More-Penrose inverse of $\mathbf{H}$.

\subsection{Dataset preprocessing and input}
\label{sec:preprocessing}

Usually, in a FNN the number of input nodes equals the number of features of the dataset. Nonetheless, our experimental scheme allows to supply the same input feature to $d$ input nodes, with $d\geq1$. We provided for this possibility since some components of ${E}^{comb}$ may be too weak to properly encode an input. For instance, in Fig.\ \ref{fig:measured_comb}, the comb component for $k=4$ is almost vanishing: if an input feature is encoded on the amplitude of this component, it will be negligible compared to other input features. Setting $d>1$ proved useful to avoid this risk, as discussed in Section \ref{sec:simulation}.

The preprocessing of the input data consists in the following operations. The input dataset is rescaled in such a way that each input feature assumes value in the range $[0,\;1]$. Then, the feature values are linearly converted into attenuations in the range $[-30 \textrm{ dB},\;0 \textrm{ dB}]$. Finally, each input entry $\mathbf{u}$ is stretched according to the selected value of $d$. If $\mathbf{u}$ contains $N$ elements, it is transformed as follows:

\begin{equation}
    \mathbf{u}=(u_1,u_2,\ldots,u_{N-1},u_N)\rightarrow(\underbrace{u_1,u_1,\ldots,u_1}_\text{$d$ times},u_2,\ldots,u_{N-1},\,\underbrace{u_N,u_N,\ldots,u_N}_\text{$d$ times}).
\end{equation}
After being preprocessed, the feature vector $\mathbf{u}$ contains the attenuations to be applied to the comb. In our experiments we always encoded the input in the most central part of the frequency comb, where most of the optical power is contained. Thus, for example, if $\mathbf{u}$ (after the stretching operation) contains $M=N\cdot d$ elements and $M$ is odd, the first attenuation $u_1$ is assigned to $F^\textrm{in}_{-(M-1)/2}$ and the last one, $u_{M}$, is assigned to $F^\textrm{in}_{(M-1)/2}$. The remaining part of the filter $\mathbf{F}^\textrm{in}$, i.e.\ the part acting on comb lines not encoding any input node, is set on zero-attenuation. Not filtering out unused parts of the input comb proved to be beneficial for tasks requiring few input nodes when operating at low $d$, most probably because this lets more power inside the system, hence leads to a richer hidden layer.

\subsection{Weight estimation}
\label{sec:readout}

A task is defined by the set of input features recorded in the matrix $\mathbf{U}$, and the set of target outputs recorded in $\mathbf{\tilde{Y}}$, which, as described before, may be a vector or a matrix according to the number of output node required. For each task, $\mathbf{U}$ is preprocessed as described in Section \ref{sec:preprocessing}, then it is split in two parts: $70\%$ of the entries constitute the "train dataset", and the remaining $30\%$ constitutes the "test dataset". The train dataset is employed to estimate the optimal set of weights $\mathbf{W}$, while the test dataset is employed to evaluate the performance of the trained network. To gather statistics about the performances, for each task we tested different random repartitions in train and test datasets.

First, the optimal set of weights $\mathbf{W}$ has to be estimated. For each input layer contained in the train dataset, the corresponding hidden layer is recorded, hence building the matrix $\mathbf{H}$. Each hidden layer node is read loading the proper notch filter, i.e.\ a filter selecting only the desired comb component, on $\textrm{SF}_2$ and redirecting its power towards one of the photodiodes. To speed-up the procedure, we exploited the dual-output capabilities of $\textrm{SF}_2$, setting two different notch filters at the same time, hence selecting two different comb lines simultaneously and redirecting them towards $\textrm{PD}_1$ and $\textrm{PD}_2$. Once $\mathbf{H}$ is recorded, the ridge regression algorithm described in Eq.\  \eqref{eq:ridge} is applied to estimate the optimal output weights $\textbf{W}$.

Once $\textbf{W}$ has been estimated, the performances of the ELM are evaluated on the train dataset, comparing the network outputs with the target ones. The output layers are obtained by multiplying the hidden layers by the output weights. This multiplication can be performed digitally or optically, as described in the following.

\paragraph{Digital weighting.}

For each entry $\mathbf{u}$ of the test dataset, the corresponding hidden layer $\mathbf{h}$ is recorded by using notch filters, as described above. Then, the output layer $\mathbf{y}=\mathbf{h}\cdot\mathbf{W}$ is calculated on the computer.

\begin{figure}[ht]
  \centering
  \includegraphics[width=.99\linewidth]{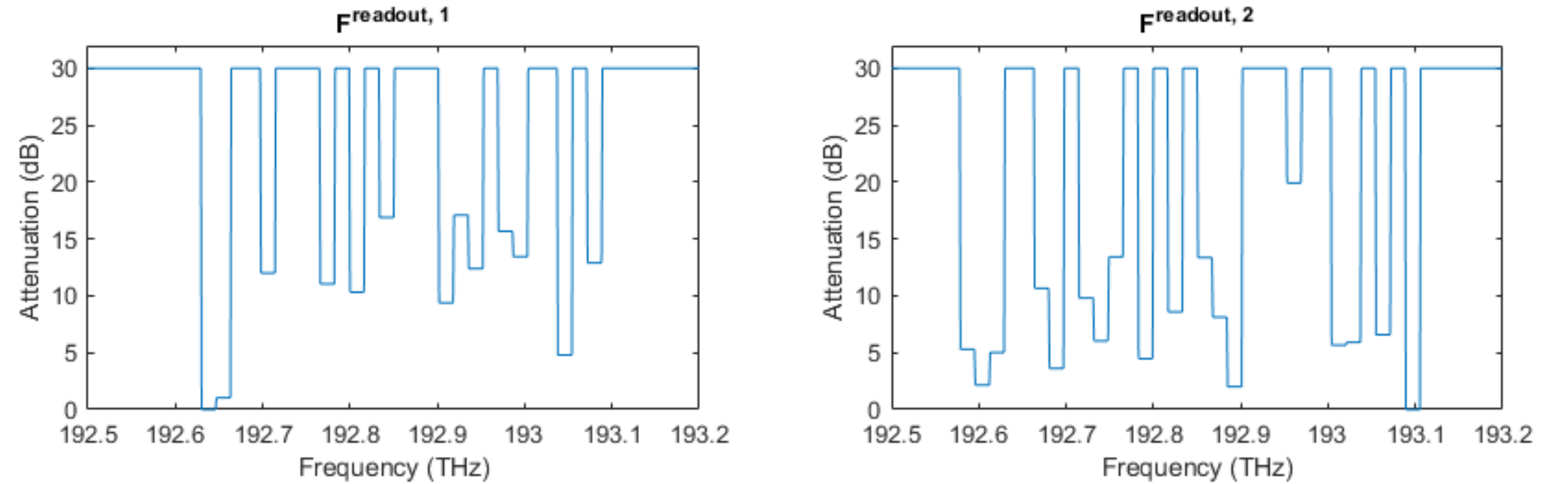}  
  \caption{Typical readout filters employed during optical weighting mode. The attenuation correspondent to complete block states is plot as $30\textrm{ db}$.}
  \label{fig:readout_weights}
\end{figure}

\paragraph{Optical weighting.}
For simplicity, first suppose that the output layer is composed of a single node, hence $\mathbf{W}$ is a column vector. Two sets of weights, $\textbf{W}^+$ and $\textbf{W}^-$, are generated from $\textbf{W}$: the first contains only the positive weights, and zeros in place of the negative one; the second contains only the negative weights, taken without sign, and zeros in place of the positive ones. 
Note that, by definition, the vectors $\textbf{W}^+$ and $\textbf{W}^-$ cannot contain two non-zero elements in the same position.
Two different filter shapes, $\textbf{F}^+$ and $\textbf{F}^-$, are then generated starting from $\textbf{W}^+$ and $\textbf{W}^-$ respectively. The procedure is similar to what employed to generate $\mathbf{F}^\textrm{in}$: the weights are rescaled in the range $[0,\;1]$ and then linearly converted into attenuations in the range $[-30\textrm{ dB},\;0\textrm{ dB}]$, with exception of the weights valued exactly zero, which are converted into a complete block state. The readout proceeds as described by Eqs. \eqref{eq:Ereadout1} and \eqref{eq:Ereadout2}, with $\textbf{F}^\textrm{readout, 1}=\textbf{F}^+$ and $\textbf{F}^\textrm{readout, 2}=\textbf{F}^-$. Figure \ref{fig:readout_weights} contains an example of readout spectral filters employed during the experiment. The result of the application of these two sets of weights are read by the two photodiodes $\textrm{PD}_1$ and $\textrm{PD}_2$. Since the photodiodes integrate power over the whole spectrum, their readings are equivalent to two linear combinations of hidden node powers, whose coefficients are the attenuations in $\textbf{F}^+$ and $\textbf{F}^-$. The output node is reconstructed as
\begin{equation}
\label{eq:opt_weights}
    y = C_+\cdot I_1+C_-\cdot I_2+C_0
\end{equation}
where $I_1$ and $I_2$ represent the readings from the two photodiodes. The set of coefficients $\textbf{C}=(C_+,\,C_-,\,C_0)$ could in principle be obtained from $\textbf{W}$. Nonetheless, we employed $10\%$ of acquired data to learn the optimal set of coefficients $\textbf{C}$ through Ordinary Least Squares algorithm. If the first $n$ entries of $\mathbf{U}$ are employed to train $\textbf{C}$, adapting Eq.\ \eqref{eq:least_squares}, the optimal set of coefficients is given by
\begin{equation}
    \textbf{C} = \begin{bmatrix} 
    C_+ \\ C_- \\ C_0
    \end{bmatrix} = \textrm{pinv}\left(\begin{bmatrix} 
    I_1^1 & I_2^1 & 1 \\
    \vdots & \vdots &\vdots \\
    I_1^n & I_2^n & 1
    \end{bmatrix}\right)\cdot\begin{bmatrix} 
    \tilde{y}^1\\ \vdots\\ \tilde{y}^n
    \end{bmatrix},
\end{equation}
where $I_{1,\,2}^i$ and $\tilde{y}^i$ are, respectively, the two intensity readings and the target output value correspondent to the i-th entry of the input dataset. Note that the column full of ones in the inverted matrix is required to learn the optimal offset $C_0$.
The set of coefficients learnt in this way performs better than the one that could be obtained from $\textbf{W}$, since in this last training phase $\textbf{C}$ is adjusted to compensate both for the presence of dark noise in the measurements and for the difference between the response of $\textrm{PD}_1$, measuring $I_1$, and $\textrm{PD}_2$, measuring $I_2$. Note that the coefficients $\mathbf{C}$ are not universal, i.e.\ they have to be calculated for each task, because they also account for the normalization of the task-dependent weights.

If the task requires an output layer composed of more than one node, the procedure here described is repeated multiple times, employing different sets $\textbf{W}^+$, $\textbf{W}^-$ and $\textbf{C}$ for each output node.

We found the optical weighting configuration to provide often better performances than the digital weighting one (see Section \ref{sec:results}). This effect is most probably due to the extra training phase introduced in optical weighing mode, as described by Eq. \eqref{eq:opt_weights}.

Finally, we point out that the optical weighting scheme does not intrinsically require a computer to perform differences: a differential amplifier is sufficient to evaluate $I_1-I_2$.

\section{Results}\label{sec:results}
We mainly tested the ELM on classification problems, such as Iris\cite{iris_dataset} and Wine\cite{wine_dataset} Classification, as well as Banknote Authentication\cite{banknote_dataset}. In classification tasks the ELM is required to assign the correct class to each "sample", i.e.\ to each set of input features. The network has as many output nodes as possible output classes, and after each readout the class corresponding to the node getting the highest value is considered to be the prediction of the network. Note that if only two classes are present, one output node is enough to encode the prediction (if $y=\mathbf{h}\cdot\mathbf{W}\leq0.5$ the network predicts the first class, otherwise it predicts the second one). Experimental results are compared both with simulations and with the scores obtained by a Support Vector Machine (SVM). We also considered the Nonlinear Channel Equalization problem\cite{jaeger2004harnessing}, which is well known in the Reservoir Computer community and is described below. The results on this task are compared both with simulation and with other experimental results in the literature.

\begin{figure}[h]
\centering
  \includegraphics[width=.6\linewidth]{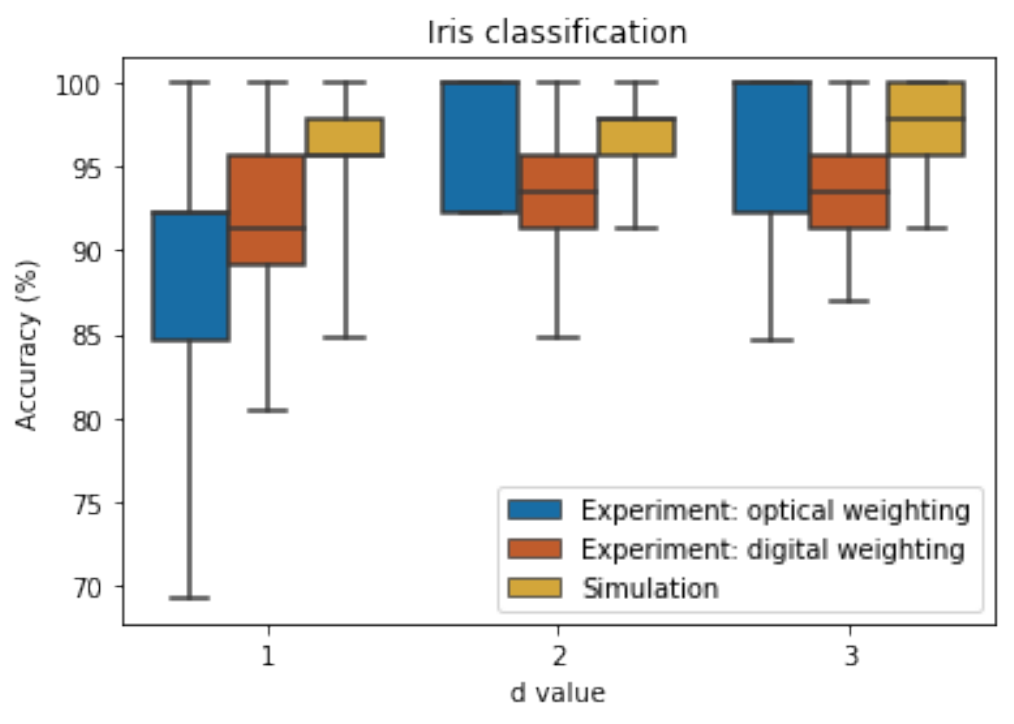}  
  \caption{Experimental and simulation result for the Iris Classification task. The boxplot diagram describes statistics obtained from $100$ cross-validation tests in the case of digital weighing and simulation, and $10$ different runs of the experiment in the case of optical weighing. The extremes of the colored boxes represent the first and third quartile of the score distributions; horizontal lines external to the boxes represent the minimum and the maximum of the score distributions; horizontal lines inside the colored boxes represent the median of the score distributions. Note that the scores are quantized, hence these elements can be superimposed. This is the case, for example, of optical weighting scores for $d=2$: the only recorded accuracies were $92.3\%$ and $100\%$ (corresponding to zero and one error respectively): hence, the minimum value equals the first quartile while the median equals the third quartile and the maximum value.}
  \label{fig:res_flower}
\end{figure}

\paragraph{Iris Classification.}
The Iris Classification task consists in selecting the correct class among three different ones, given a set of four different features. The ELM is thus composed of 4 input nodes and 3 output nodes, one for each possible output class. Performances on the Iris Classification task are reported in Figure \ref{fig:res_flower}. In digital weighting mode the ELM reached an accuracy of $93.9 \%$ (setting $d=3$ and $\lambda=10^{-7}$). The average accuracy recorded over 10 optical weighting runs was $97.7 \%$ (setting $d=2$). A Support Vector Machine reached an accuracy of $98.0\%$.

\begin{figure}[h]
\centering
  \includegraphics[width=.6\linewidth]{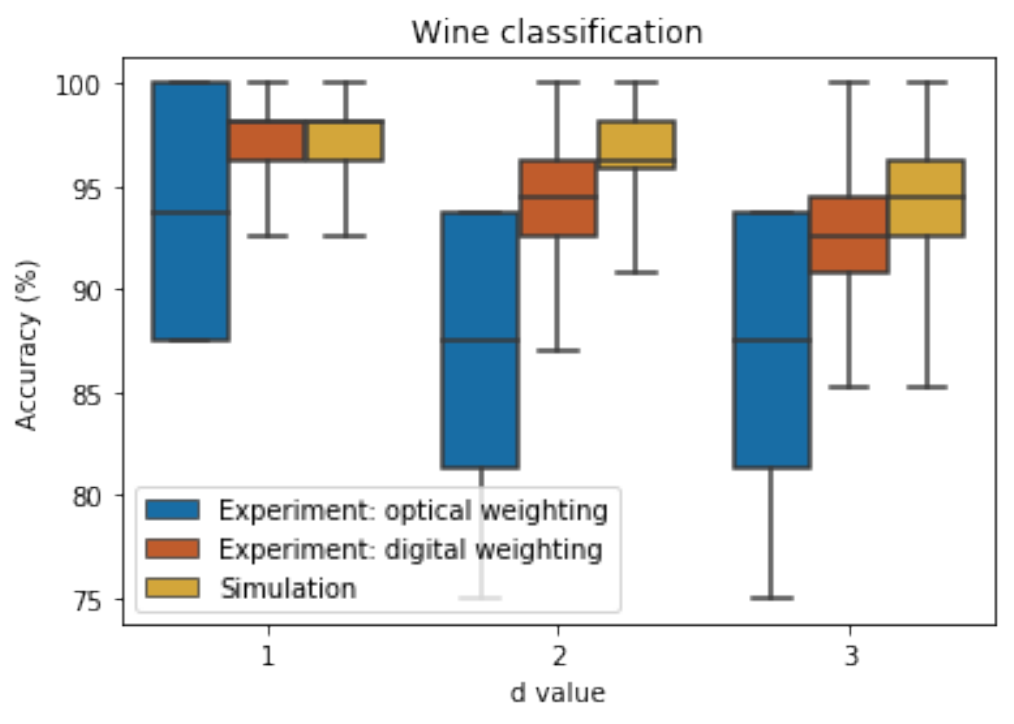}  
  \caption{Experimental and simulation result for the Wine Classification task. The boxplot diagram (see Figure \ref{fig:res_flower}) describes statistics obtained from $100$ cross-validation tests in the case of digital weighing and simulation, and $10$ different runs of the experiment in the case of optical weighing.}
  \label{fig:res_wine}
\end{figure}

\paragraph{Wine Classification.}
The Wine Classification task consists in selecting the correct class among three different ones, given a set of thirteen different features. The ELM is thus composed of 13 input nodes and 3 output nodes, one for each possible output class. Performances on the Wine Classification task are reported in Figure \ref{fig:res_wine}. In digital weighting mode the ELM reached an accuracy of $97.5 \%$ (setting $d=1$ and $\lambda=10^{-6}$). The average accuracy recorded over 10 optical weighting runs was $94.4 \%$ (setting $d=1$). A Support Vector Machine reached an accuracy of $97.8\%$.

\begin{figure}[h]
\centering
  \includegraphics[width=.6\linewidth]{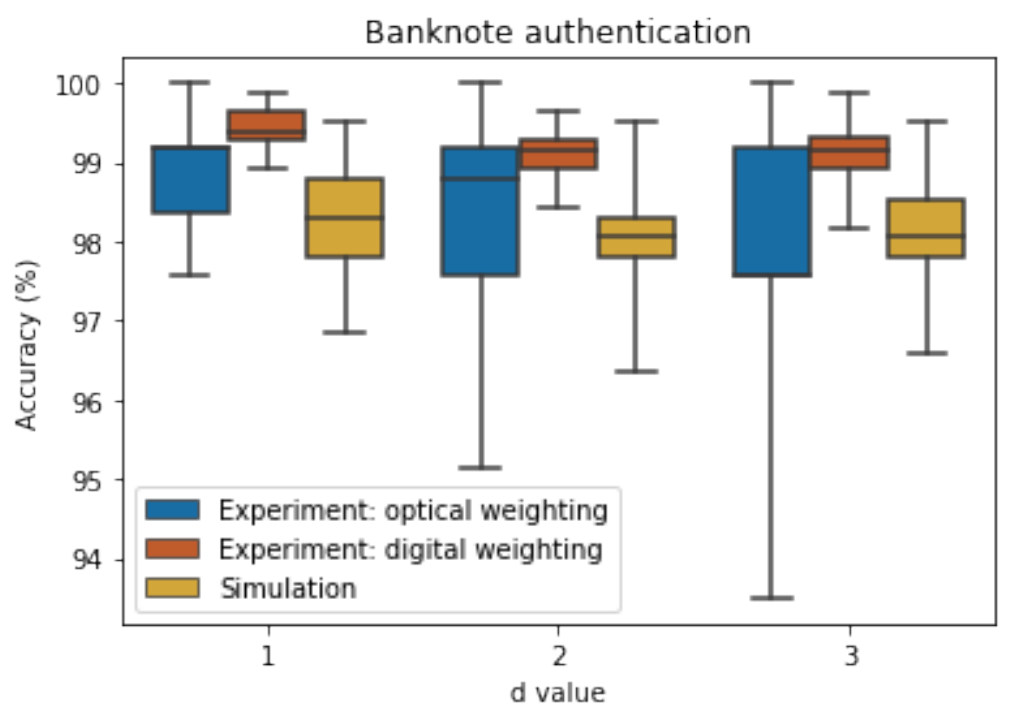}  
  \caption{Experimental and simulation result for the Banknote Authentication task. The boxplot diagram (see Figure \ref{fig:res_flower}) describes statistics obtained from $100$ cross-validation tests in the case of digital weighing and simulation, and $10$ different runs of the experiment in the case of optical weighing.}
  \label{fig:res_banknote}
\end{figure}

\paragraph{Banknote Classification.}
The Banknote Classification task consists in selecting the correct class among two different ones, given a set of five different features. The ELM is thus composed of 5 input nodes and 1 output node, which is enough to encode the two possible classes. Performances on the Banknote Authentication task are reported in Figure \ref{fig:res_banknote}. In digital weighting mode the ELM reached an accuracy of $99.4 \%$ (setting $d=1$ and $\lambda=10^{-5}$). The average accuracy recorded over 10 optical weighting runs was $98.8 \%$ (setting $d=1$). A Support Vector Machine reached an accuracy of $100\%$.

\begin{figure}[h]
\centering
  \includegraphics[width=.75\linewidth]{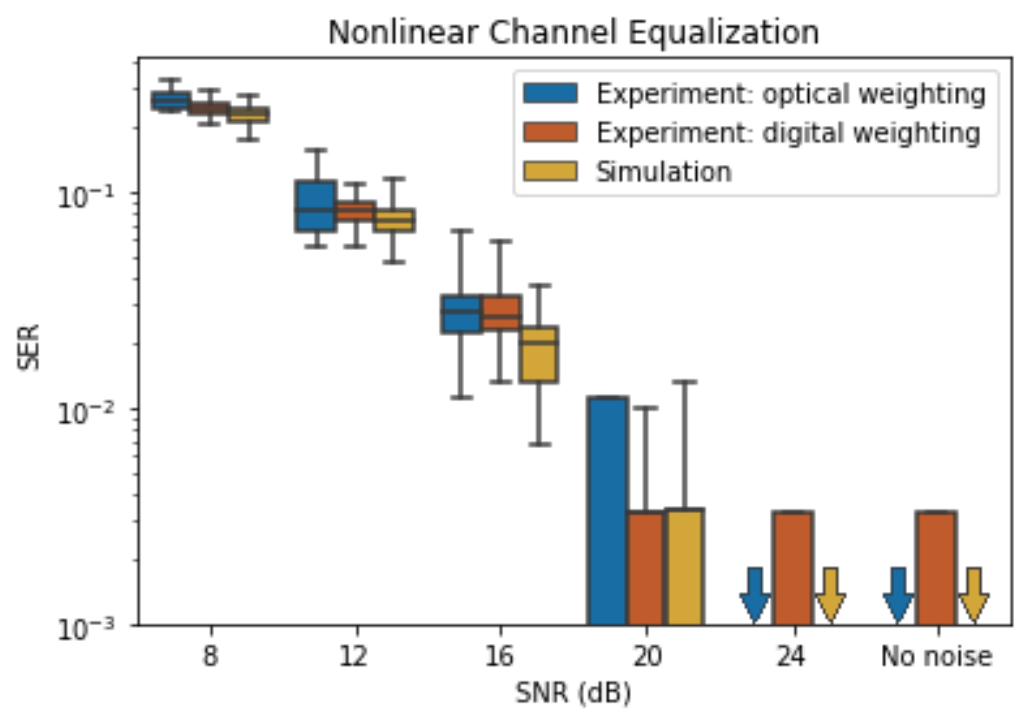}  
  \caption{Experimental and simulation result for the Nonlinear Channel Equalization task. All the experiments are executed with $d=2$ and $\lambda=10^{-9}$ over $1000$ transmitted symbols. The boxplot diagram (see Figure \ref{fig:res_flower}) describes statistics obtained from $100$ cross-validation tests in the case of digital weighing and simulation, and $10$ different runs of the experiment in the case of optical weighing. Downward pointing arrows indicate that no errors have been recorded after $1000$ transmissions, hence $\textrm{SER}<10^{-3}$. Note that in two less noisy configurations ($\textrm{SNR}=24\textrm{ dB}$ and no-noise) digital weighting always recorded either no error or only one error per run.}
  \label{fig:res_NLC}
\end{figure}

\paragraph{Nonlinear Channel Equalization.}
The Nonlinear Channel Equalization task consists in reconstructing a signal after the transmission through a channel which induces a nonlinear distortion and has memory. The input signal is a sequence of random symbols $u(t)$ uniformly extracted from the set $\{-3,\,-1,\,1,\,3\}$. This signal first goes through a linear channel exhibiting memory effects: 
\begin{equation*}
    \begin{split}
        q(t) &= 0.08u(t+2)-0.12u(t+1)+u(t)+0.18u(t-1)\\
        &-0.1u(t-2)+0.091u(t-3)-0.05u(t-4)\\
        &+0.04u(t-5)+0.03u(t-6)+0.01u(t-7),
    \end{split}
\end{equation*}
and then through a noisy nonlinear channel: \begin{equation*}
    x(t)=q(t)+0.036q(t)^2-0.011q(t)^3+\nu(t),
\end{equation*}
where $\nu(t)$ is a Gaussian noise with a power selected in such a way to achieve a certain desired Signal to Noise Ratio (SNR). For each timestep $t$, the channel outputs $x(t-7)$, $x(t-6)$, $...$, $x(t+1)$, $x(t+2)$ are supplied to the ELM and the task consists in reconstructing $u(t)$. Thus, this task is equivalent to a classification in four different possible classes given ten input features. Contrary from previous tasks, here we employ only one output node and we take as output of the ELM the value in the set $\{-3,\,-1,\,1,\,3\}$ closest to the output node value. We tested the performances over different SNR values, ranging from $8\textrm{ dB}$ to $24\textrm{ dB}$ and in a no-noise configuration. Performances of the Nonlinear Channel Equalization task are evaluated by the Symbol Error Rate (SER), i.e.\ the ratio between errors and total transmitted symbols, and are reported in Figure \ref{fig:res_NLC}. These results are obtained setting $d=2$ and the best performing $\lambda$ value for each SNR (selected $\lambda$ values belong to the range $[10^{-10},\,10^{-5}]$). In terms of SER, our ELM running in optical weighting outperforms by almost one order of magnitude a previous optical implementations of a time-multiplexed ELMs \cite{ortin2015unified}.\footnote{The ELM described in \cite{ortin2015unified} receives as input a set of channel states in $7$ different times and employs $247$ hidden nodes featuring a $\sin^2$ nonlinearity; our ELM receives as input a set of channel states in $10$ different times and employs $31$ linear hidden nodes, the only nonlinearity being introduced in the readout.} Our ELM was tested on 1000 symbols, hence SERs less than $10^{-3}$ are undetectable. Increasing the order of magnitude of the input symbols count is currently experimentally unfeasible, due to the slow settling time of the programmable filters. However, in numerical simulation we found SERs of $2.3\cdot10^{-4}$ for an SNR of $28\textrm{ dB}$, $7.2\cdot10^{-5}$ for an SNR of $32\textrm{ dB}$ and $2.8\cdot10^{-5}$ in a no-noise configuration. These simulated performances are comparable with the ones obtained by Reservoir Computers (RC) reported in literature \cite{vinckier2015high, paquot2012optoelectronic, duport2012all, dejonckheere2014all}, and, in some cases, even almost one order of magnitude better. Note that these Reservoir Computing approaches also rely on the capability of the network to memorize previous input, since only the current state of the channel is supplied as input in each timestep. Contrary to RCs, an ELM does not have memory of the past inputs, since the network features no recurrency. As a consequence, for the Channel Equalization task, memory has to be implemented outside the network: both in our case and in \cite{ortin2015unified} it is implemented in the script generating input layers, as described above.

\subsection{Dependence on hyperparameters}\label{sec:simulation}
\begin{figure}[ht]
\centering
\includegraphics[width=.7\linewidth]{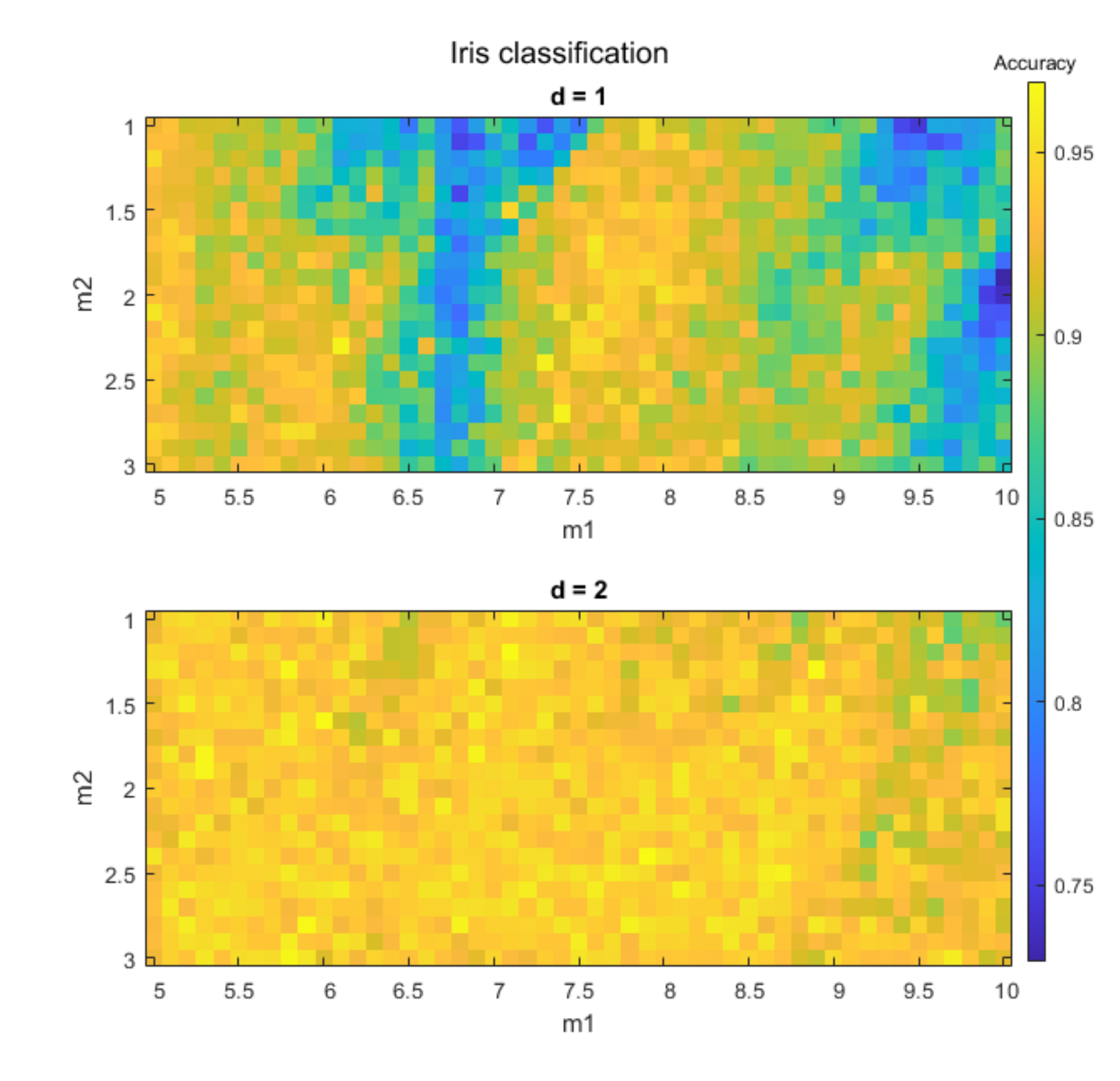}  
\caption{Simulated accuracy for Iris Classification task with $d=1$ and $d=2$ as a function of $m_1$ and $m_2$. Higher is better.}
\label{fig:performances_iris}
\end{figure}

\begin{figure}[ht]
\centering
\includegraphics[width=.7\linewidth]{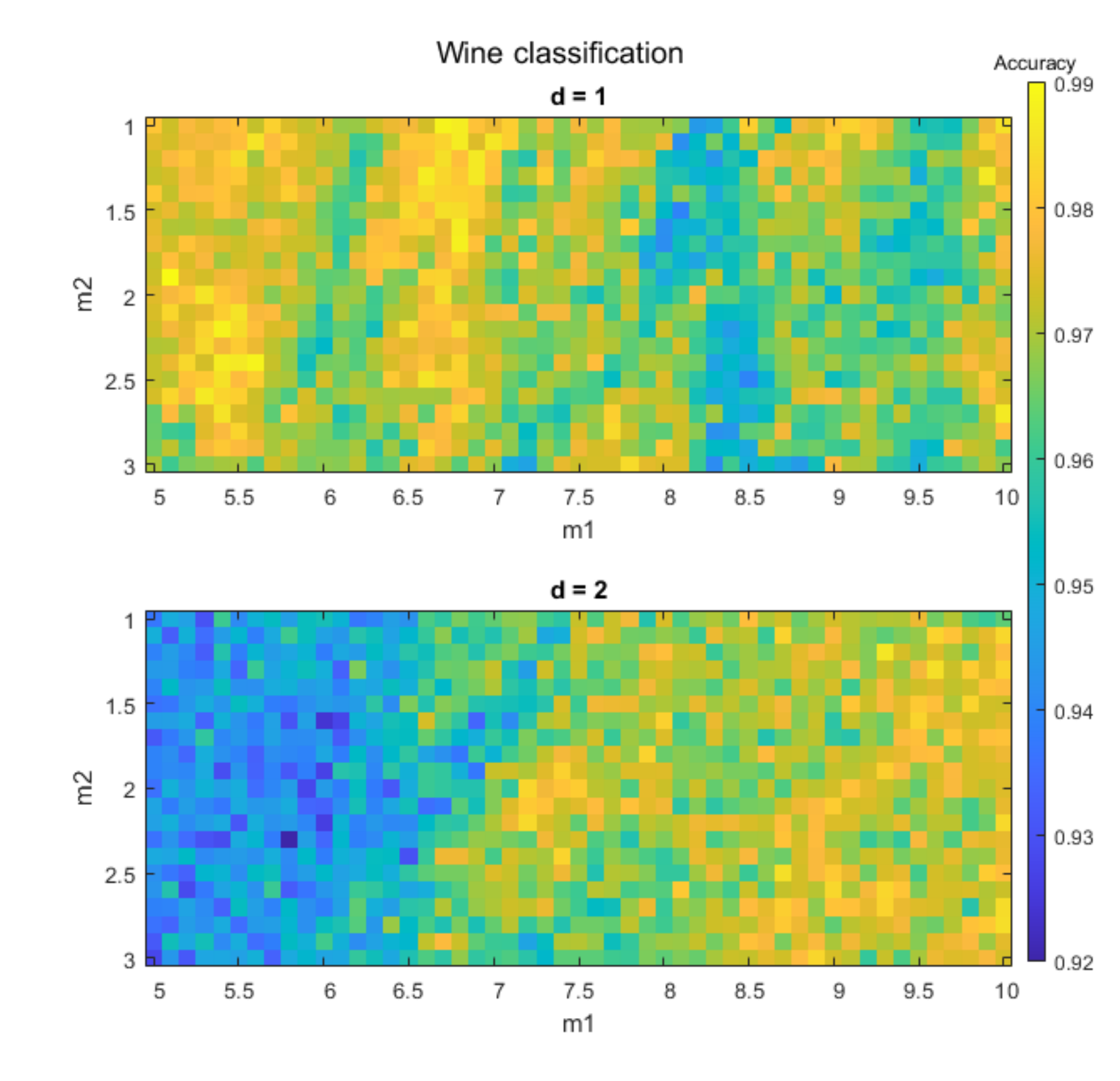} 
\caption{Simulated accuracy for Wines Classification task with $d=1$ and $d=2$ as a function of $m_1$ and $m_2$. Higher is better.}
\label{fig:performances_wine}
\end{figure}

\begin{figure}[ht]
\centering
\includegraphics[width=.7\linewidth]{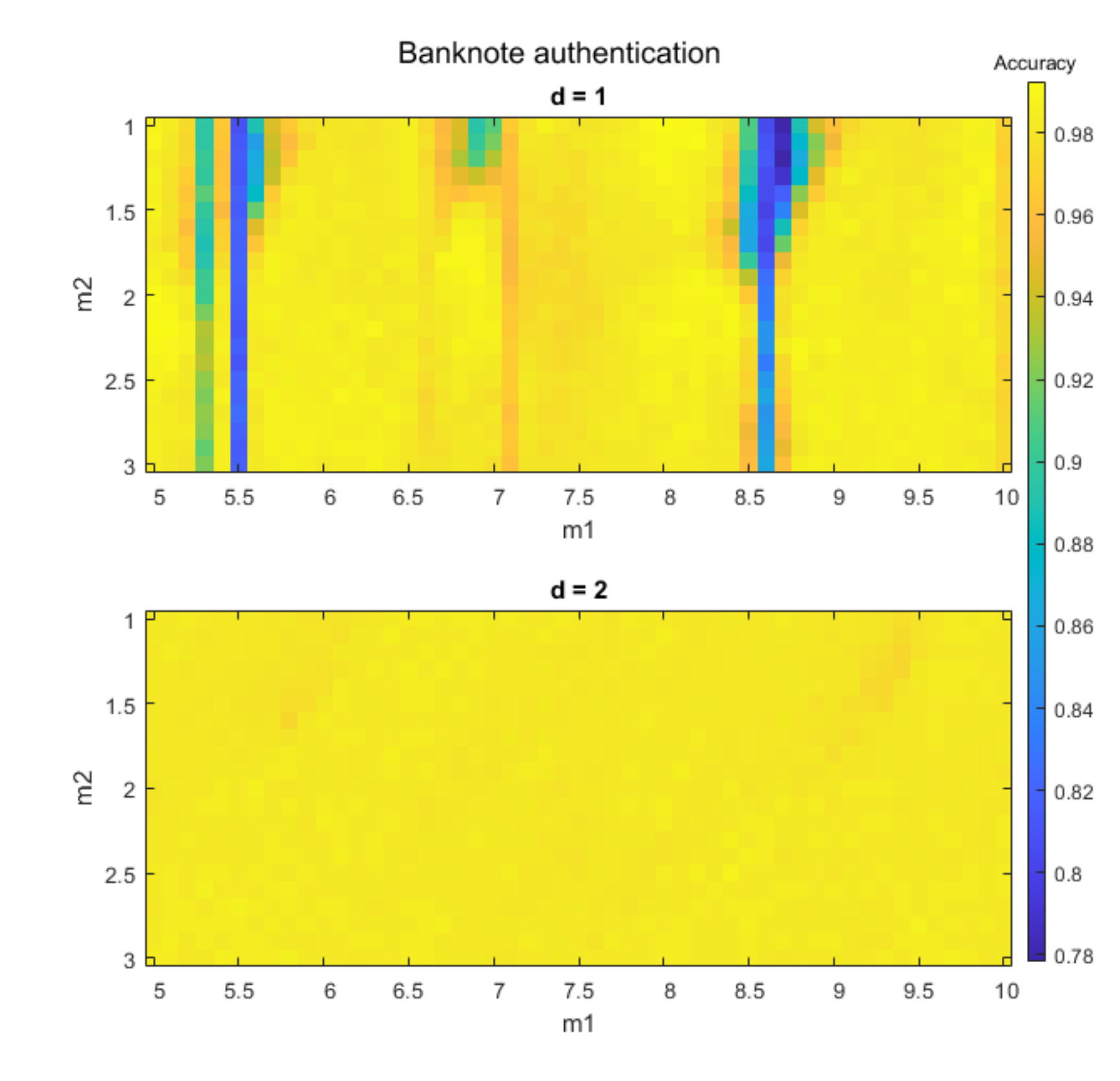}  
\caption{Simulated accuracy for Banknote Authentication task with $d=1$ and $d=2$ as a function of $m_1$ and $m_2$. Higher is better. Sharp drops in performance can be noted in this plot, when $d=1$ and $m_1\approx5.3$ or $m_1\approx8.6$.}
\label{fig:performances_banknote}
\end{figure}

\begin{figure}[ht]
\centering
\includegraphics[width=.7\linewidth]{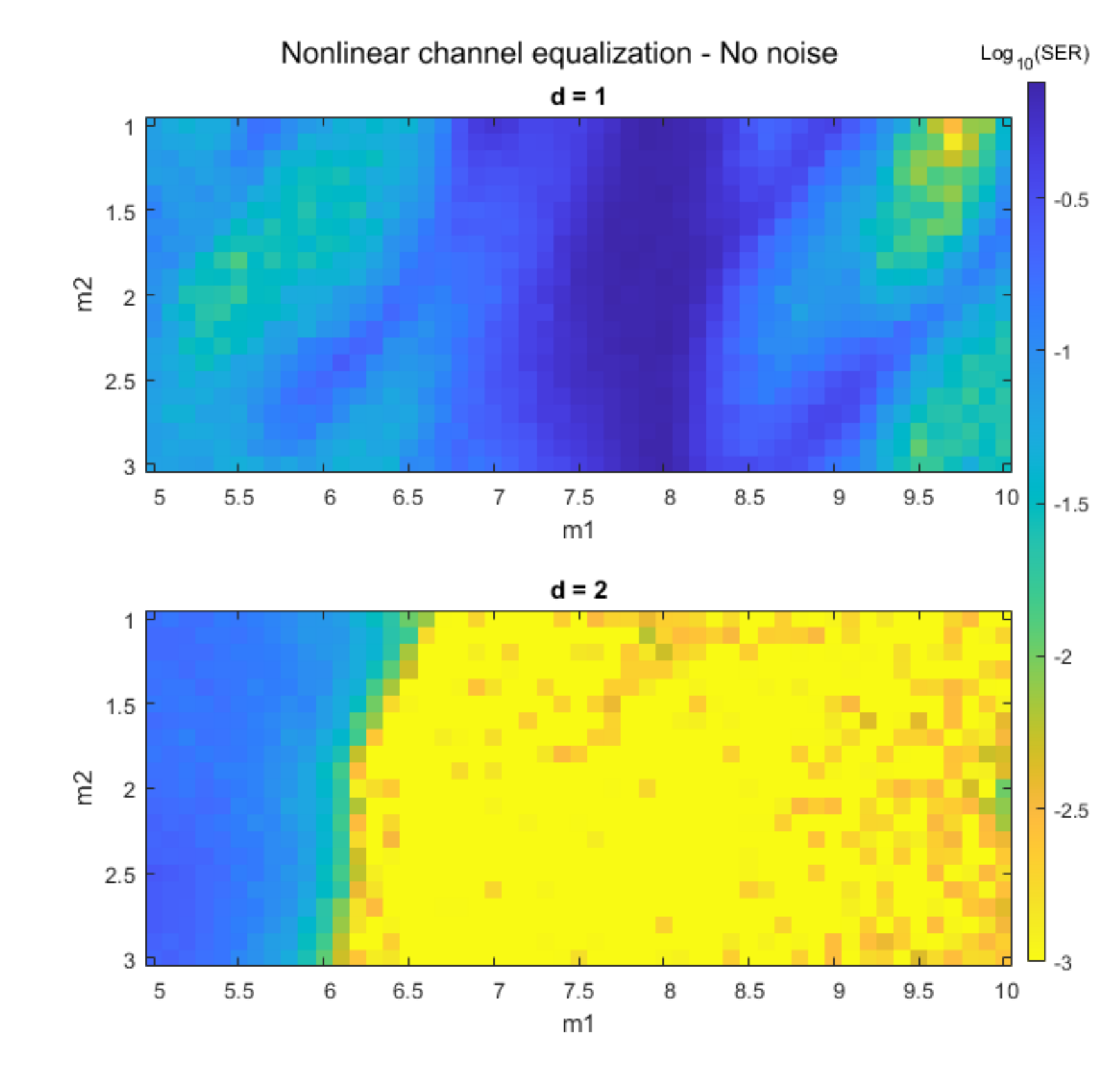} 
\caption{Simulated Symbol Error Rate for Nonlinear Channel Equalization task without noise and with $d=1$ and $d=2$ as a function of $m_1$ and $m_2$. Logarithmic scale, low is better. To improve readability of the color map, all SERs less than $10^{-3}$, including many zeroes, are plotted as  $10^{-3}$.}
\label{fig:performances_nlc60}
\end{figure}

\begin{figure}[ht]
\centering
\includegraphics[width=.7\linewidth]{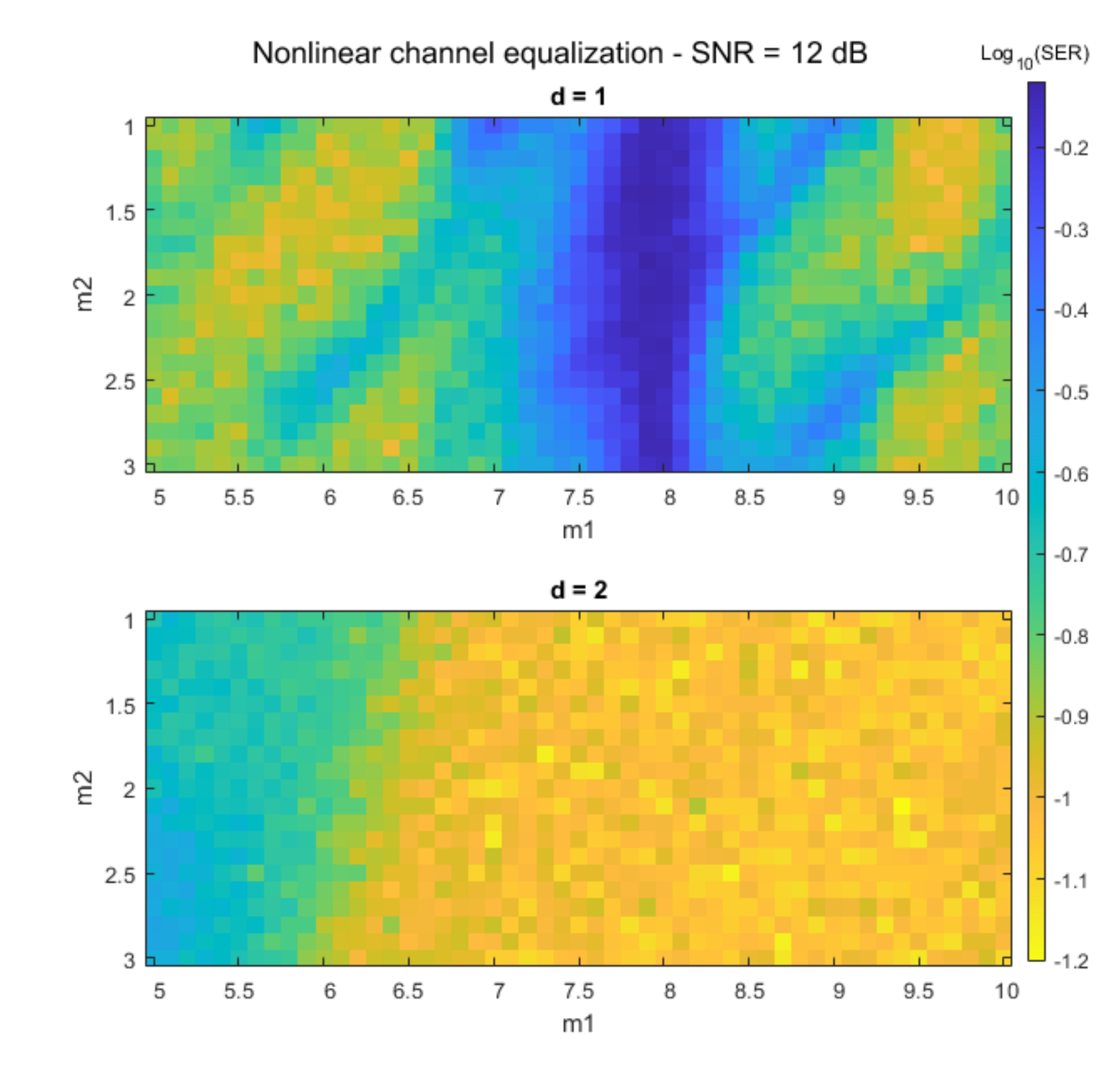} 
\caption{Simulated Symbol Error Rate for Nonlinear Channel Equalization task with $SNR=12\textrm{ dB}$ and with $d=1$ and $d=2$ as a function of $m_1$ and $m_2$. Logarithmic scale, low is better. Note the sharp drop in performances when $d=1$ and $m_1\approx7.9$, due to the particular shape of the input comb, unable to encode an important feature.}
\label{fig:performances_nlc12}
\end{figure}

The system has been simulated according to the model described in Section \ref{sec:electric_field}. For each input layer, the corresponding hidden layer is simulated and the output layer is calculated as described in Section \ref{sec:readout} in the 'digital readout' case. The simulation allows to evaluate performances systematically scanning the hyperparameters $d$, $m_1$ and $m_2$. Note that such an accurate scan is unfeasible in the experimental setup, both because of the prohibitive time it would require and for the impossibility of setting $m_1$ and $m_2$ independently. We scanned the performances of each tested task: Iris (Figure \ref{fig:performances_iris}) and Wine classification (Figure \ref{fig:performances_wine}), Banknote authentication (Figure \ref{fig:performances_banknote}) and NLC (Figures \ref{fig:performances_nlc60} and \ref{fig:performances_nlc12}).

The simulated scans allow two observations about the working mechanism of this ELM. First, when $d=1$, the performances are extremely dependent on $m_1$ and show sharp drops for certain values of this hyperparameter. We found that the positions of these drops depend on the arrangement of the input features. Since this effect is strongest when $d=1$, we conclude that drops in performance happen when an important input feature is assigned to a comb component too weak to encode it properly. Second, when $d=2$, Nonlinear Channel Equalization and Wine classifier perform badly for $m_1$ values too low. These two tasks require many features, respectively 10 and 13, which, when $d=2$, are encoded in 20 and 26 input nodes respectively. Hence, they can be completely encoded only when the input comb is large enough, that is when $m_1$ is large enough. This last effect also applies to all the other tasks when $d=3$. Scans of the performances when $d=3$ do not display any additional interesting feature and are not reported here.

\begin{figure}[ht]
\begin{subfigure}{.5\textwidth}
  \centering
  \includegraphics[width=.99\linewidth]{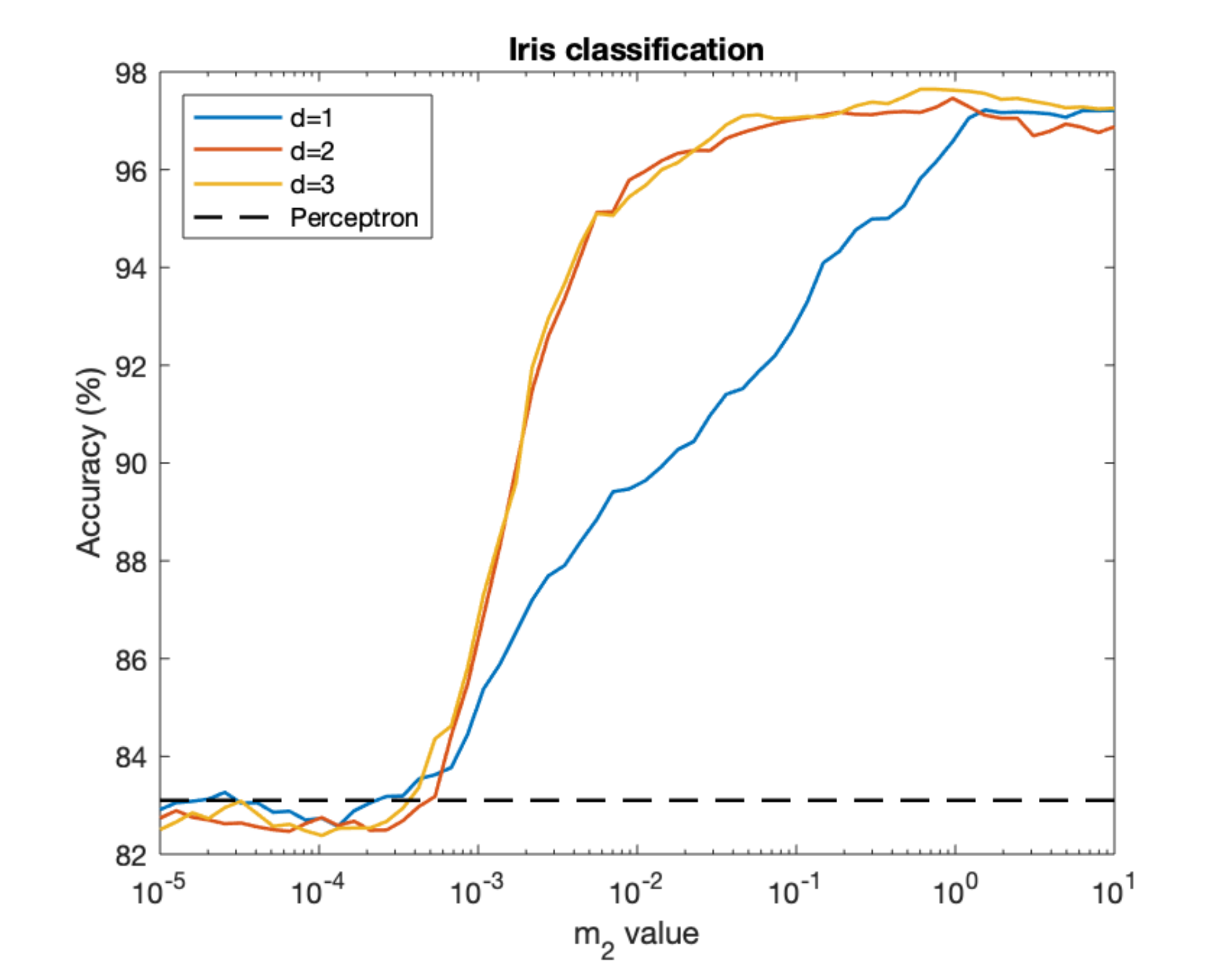}  
  \caption{}
  \label{fig:low_m2_banknote}
\end{subfigure}
\begin{subfigure}{.5\textwidth}
  \centering
  \includegraphics[width=.99\linewidth]{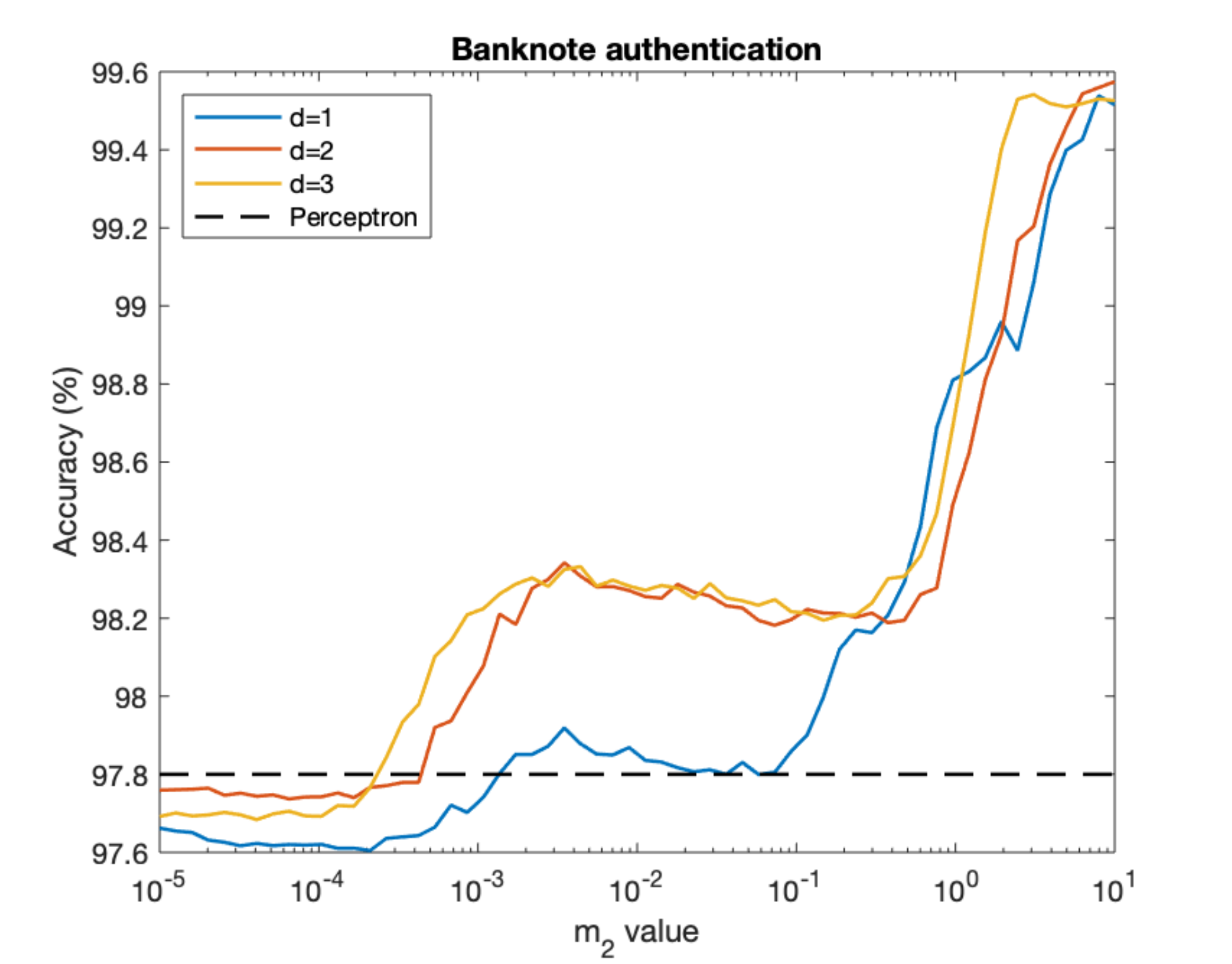}  
  \caption{}
  \label{fig:low_m2_iris}
\end{subfigure}
\caption{Simulated accuracy varying $m_2$ with $m_1=7.87$ for the Iris classification (a) and the banknote authentication (b) tasks. When $m_2$ is too small, the mixing effect provided by the Phase Modulator $\textrm{PM}_2$ is negligible and the hidden layer is identical to the input one, thus the accuracies are comparable to the ones obtained by a perceptron.}
\label{fig:low_m2}
\end{figure}

Simulation scans also suggest that a high $m_2$ parameter is not a prerequisite for good performances. In Fig.\ \ref{fig:low_m2} we plot the simulated accuracy versus $m_2$, keeping $m_1$ equal to the experimental value of $7.87$, for two selected tasks which are the most sensitive to $m_2$ variations. $m_2$ determines how strongly input nodes are mixed to generate the hidden layer, hence when $m_2=0$ the hidden layer is an exact copy of the input one. We checked that in this situation the network performs similar to a perceptron, i.e.\ a machine learning algorithm whose output is simply a linear combination of input features.

Also the arrangement of input features is a free parameter. As described in Section \ref{sec:preprocessing}, during the experiments we encoded the inputs in the central part of the comb, assigning each feature to $d$ consecutive comb lines. However, alternative schemes could be employed: for example, the same feature could be assigned to $d$ non-consecutive lines, or features could be encoded in the most powerful lines of the comb, regardless of their position. Numerical simulations suggest that these approaches do not affect performances sensibly, but they could be investigated more in the future.

\section{Conclusion}
\label{sec:conclusion}
An Extreme Learning Machine consists in a randomly initialized Feed-Forward Neural network where only output connections are trained. This concept can be translated from software to real physical substrates, exploiting the transformation that a certain system acts between its input and output spaces.
We demonstrated the feasibility of an ELM implemented in a frequency-multiplexing optical fiber setup, where also multiplication by output weights can be performed optically. Our experiment can be interpreted as an interferometer in the frequency domain, and is very stable: weights learned one day can be used the day after with no recalibration. 

The current scheme is affected by two main limitations. The first consists in the speed of execution of the experiment. This is currently limited by the programmable filters settling time, which is $\sim500\textrm{ ms}$. We expect to be possible to achieve an update rate at least comparable to the video frequency of $60\textrm{ Hz}$ by employing LCD-based optical filters. The second limit consists in the topology of the network. The number of input nodes could be increased by increasing the power of the RF signal applied on $\textrm{PM}_1$. However, the strength of the mixing, i.e.\ the number of input nodes contributing to the state of a hidden node, depends only on the power of the RF signal applied to $\textrm{PM}_2$.

Typical parallelization potentialities offered by the optical field remain to be tested. As example, more than one input wavelength could lead to improvements in the scheme: one could have multiple superimposing or not-superimposing combs, which could enrich the dynamics, increase the size of input and hidden layers, or even allow for the execution in parallel of multiple tasks. This will be studied both numerically and experimentally in the future.

\section*{Acknowledgements}
The authors thank Elger Vlieg for his contribution to the conception of this experiment and thank Ghent University - IMEC for loan of a Waveshaper.

\section*{Founding}
The authors acknowledge financial support of the European Union through the Marie Skłodowska-Curie Innovative Training Networks action POST-DIGITAL project number 860830, and from the Fonds de la Recherche Scientifique (FRS-FNRS).

\section*{Disclosures}
The authors declare no conflicts of interest.

\printbibliography

\end{document}